\documentclass{aa}
\pdfoutput=1

\usepackage{graphicx}
\usepackage{subfig}
\usepackage[varg]{txfonts}
\usepackage{lscape}
\usepackage{color}
\usepackage{longtable}

\newcommand{\mearth}{M_\oplus}

\def\ms{\hbox{\,m\,s$^{-1}$}}         
\def\m2s2{\hbox{\,m$^{2}$\,s$^{-2}$}} 
\def\sini{\hbox{sin\,$i$}}      

\begin{document}

\title{The HADES RV Programme with HARPS-N@TNG\thanks{Based on: observations made with the Italian \textit{Telescopio Nazionale Galileo} (TNG), operated on the island of La Palma by the INAF - \textit{Fundaci\'on Galileo Galilei} at the \textit{Roche de Los Muchachos} Observatory of the \textit{Instituto de Astrof\'isica de Canarias} (IAC); photometric observations made with the robotic telescope APT2 (within the EXORAP program) located at Serra La Nave on Mt. Etna.}}
\subtitle{VIII. Gl15A: A multiple wide planetary system sculpted by binary interaction}
\titlerunning{Gl15A}

\author{M. Pinamonti\inst{\ref{inst3}}
       \and
       M. Damasso\inst{\ref{inst3}}
       \and
       F. Marzari\inst{\ref{inst4}}
       \and
       A. Sozzetti\inst{\ref{inst3}}
       \and
       S. Desidera\inst{\ref{inst5}}
       \and
       J. Maldonado\inst{\ref{inst6}}
       \and
       G. Scandariato\inst{\ref{inst7}}
       \and
       L. Affer\inst{\ref{inst6}}
       \and
       A. F. Lanza\inst{\ref{inst7}}
       \and
       A. Bignamini\inst{\ref{inst2}}
       \and
       A. S. Bonomo\inst{\ref{inst3}}
       \and
       F. Borsa\inst{\ref{inst10}}
       \and
       R. Claudi\inst{\ref{inst5}}
       \and
       R. Cosentino\inst{\ref{inst7},\ref{inst8}}
       \and
       P. Giacobbe\inst{\ref{inst3}}
       \and
       E. Gonz\'alez-\'Alvarez\inst{\ref{inst6},\ref{inst9}}
       \and
       J. I. Gonz\'alez Hern\'andez\inst{\ref{inst11},\ref{inst12}}
       \and
       R. Gratton\inst{\ref{inst5}}
       \and
       G. Leto\inst{\ref{inst7}}
       \and
       L. Malavolta\inst{\ref{inst4},\ref{inst5}}
       \and
       A. Martinez Fiorenzano\inst{\ref{inst8}}
       \and
       G. Micela\inst{\ref{inst6}}
       \and
       E. Molinari\inst{\ref{inst8}}
       \and
       I. Pagano\inst{\ref{inst7}}
       \and
       M. Pedani\inst{\ref{inst8}}
       \and
       M. Perger\inst{\ref{inst13},\ref{inst15}}
       \and
       G. Piotto\inst{\ref{inst4},\ref{inst5}}
       \and
       R. Rebolo\inst{\ref{inst11},\ref{inst12}}
       \and
       I. Ribas\inst{\ref{inst13},\ref{inst15}}
       \and
       A. Su\'arez Mascare\~no\inst{\ref{inst11},\ref{inst14}}
       \and
       B. Toledo-Padr\'on\inst{\ref{inst11},\ref{inst12}}}

\institute{INAF - Osservatorio Astrofisico di Torino, Via Osservatorio 20, I-10025 Pino Torinese, Italy\label{inst3}
      \and
          Dipartimento di Fisica e Astronomia, Universit\`a di Padova, via Marzolo 8, I-35131 Padova, Italy\label{inst4}
      \and
          INAF - Osservatorio Astronomico di Padova, vicolo dell'Osservatorio 5, I-35122 Padova, Italy\label{inst5}
      \and
          INAF - Osservatorio Astronomico di Palermo, piazza del Parlamento 1, I-90134 Palermo, Italy\label{inst6}
      \and
          INAF - Osservatorio Astrofisico di Catania, Via S. Sofia 78, I-95123 Catania, Italy\label{inst7}
      \and
          INAF - Osservatorio Astronomico di Trieste, via G. B. Tiepolo 11, I-34143 Trieste, Italy\label{inst2}
      \and
          Fundaci\'on Galileo Galilei - INAF, Ramble Jos\'e Ana Fernandez P\'erez 7, E-38712 Bre\~na Baja, TF, Spain\label{inst8}
      \and
          Dipartimento di Fisica e Chimica, Universit\`a di Palermo, piazza del Parlamento 1, I-90134 Palermo, Italy\label{inst9}
      \and
          INAF - Osservatorio Astronomico di Brera, via E. Bianchi 46, I-23807 Merate, Italy\label{inst10}
      \and
          Instituto de Astrof\'isica de Canarias (IAC), E-38205 La Laguna, Tenerife, Spain\label{inst11}
      \and
          Universidad de La Laguna, Dpto. Astrof\'isica, E-38206 La Laguna, Tenerife, Spain\label{inst12}
      \and
          Institut de Ci\`encies de l'Espai (ICE, CSIC), Campus UAB, C/ de Can Magrans s/n, E-08193 Cerdanyola del Vall\`es, Spain\label{inst13}
      \and
          Observatoire Astronomique de l'Universit\'e de Gen\'eve, 1290 Versoix, Switzerland\label{inst14}
      \and
          Institut d'Estudis Espacials de Catalunya (IEEC), C/ Gran Capit\`a 2-4, E-08034 Barcelona, Spain\label{inst15}}
             
\date{}

\abstract{We present 20 years of radial velocity (RV) measurements of the M1 dwarf Gl15A, combining 5 years of intensive RV monitoring with the HARPS-N spectrograph   
with 15 years of archival HIRES/Keck RV data. We carry out an MCMC-based analysis of the RV time series, inclusive of Gaussian Process (GP) approach to the description 
of stellar activity induced RV variations. 
   
Our analysis confirms the Keplerian nature and refines the orbital solution for the 11.44-day period super Earth, Gl15A\,b, reducing its amplitude to
$1.68^{+0.17}_{-0.18}$ m s$^{-1}$ ($M \sin i = 3.03^{+0.46}_{-0.44}$ $\mearth$), and successfully models a long-term trend in the combined RV dataset in terms of a
Keplerian orbit with a period around 7600 days and an amplitude of $2.5^{+1.3}_{-1.0}$ m s$^{-1}$, corresponding to a super-Neptune mass ($M \sin i = 36^{+25}_{-18}$
$\mearth$) planetary companion.
   
We also discuss the present orbital configuration of Gl15A planetary system in terms of the possible outcomes of Lidov-Kozai interactions with the wide-separation 
companion Gl15B in a suite of detailed numerical simulations. In order to improve the results of the dynamical analysis, we derive a new orbital solution for the binary 
system, combining our RV measurements with astrometric data from the WDS catalogue.
  
The eccentric Lidov-Kozai analysis shows the strong influence of Gl15B on the Gl15A planetary system, which can produce orbits compatible with the observed 
configuration for initial inclinations of the planetary system between $75^\circ$ and $90^\circ$, and can also enhance the eccentricity of the outer planet well above 
the observed value, even resulting in orbital instability, for inclinations around $0^\circ$ and $15^\circ - 30^\circ$.
   
The Gl15A system is the multi-planet system closest to Earth, at $3.57$ pc, and hosts the longest period RV sub-jovian mass planet discovered so far. Its orbital 
architecture constitutes a very important laboratory for the investigation of formation and orbital evolution scenarios for planetary systems in binary stellar 
systems.}
   

\maketitle
   
%

\section{Introduction}

   Extrasolar planetary systems always showed a huge variety of orbital architectures, ever since the very first exoplanet orbiting a main sequence star was discovered \citep{mayorqueloz95}. In the following two decades the number of known exoplanets grew up to more than 3500\footnote{\url{https://exoplanetarchive.ipac.caltech.edu/} - 18/09/2017}, mainly discovered from transit (e.g., Kepler) and radial-velocity (e.g, HARPS, HARPS-N) observations, over a wide spread range of masses, radii and orbital separations, aiming down to the identification of small mass rocky Earth-twins \citep[e.g.][]{pepeetal11}.
   
   Low mass M dwarfs are the most common main sequence stars, comprising $\sim 70 \%$ of the stars in the Solar Neighbourhood \citep{henryetal06}. Furthermore, M dwarfs have become the most promising ground for the hunt for low-mass, rocky planets \citep[e.g.][]{dreschar2013,sozzettietal13,astudillodefruetal2017}, due to their more advantageous mass and radius ratios compared to solar-type stars.
   
   There is a solid evidence, arising both from HARPS and Kepler observations, that super Earths and Neptunes are commonly found in multiple systems \citep[e.g.][and references therein]{udryetal2017,roweetal2014}.

   The nearby M1 dwarf \object{Gl15A} was studied by \citet{howardetal2014}, who found a short-period super-Earth orbiting the star: they measured a period of $11.44$ d and an amplitude of $2.94$ m s$^{-1}$. They also studied the activity signals of the host star, identifying the rotational period of the star to be $44$ d, both from the $S_\text{HK}$ index analysis and also from the precise photometric light-curve, collected with the automatic photometric telescope (APT) at Fairborn Observatory \citep{eatonetal2003}.
   Gl15A has also a known M3.5 binary companion, Gl15B, identified astrometrically by \citet{lippincott1972} from a small fragment of its orbit, with a measured orbital separation of $146$ AU, and an orbital period of $2600$ yr.
   
   In this framework we present the clear detection of a long-period eccentric super-Neptune planet around \object{Gl15A}, from 5 years of high-precision Doppler monitoring with the HARPS-N high-resolution (resolving power $R \sim 115000$) optical echelle spectrograph \citep{cosentinoetal2012} at the Telescopio Nazionale Galileo (TNG), combined with 15 years of archival RV data from the LCES HIRES/Keck Precision Radial Velocity survey \citep{butleretal2017}. The HARPS-N data were collected as part of the HADES (Harps-n red Dwarf Exoplanet Survey) programme, a collaboration between the Italian GAPS \citep[Global Architecture of Planetary Systems,][]{covinoetal2013,desideraetal2013,porettietal2016} Consortium\footnote{\url{http://www.oact.inaf.it/exoit/EXO-IT/Projects/Entries/2011/12/27_GAPS.html}}, the Institut de Ci\`encies de l'Espai de Catalunya (ICE) and the Instituto de Astrof\'isica de Canarias (IAC). We also confirm the presence and update the amplitude of the Gl15A\,b RV signal, significantly reducing its minimum mass, while confirming the orbital period measured by \citet{howardetal2014}.
   Our findings are also discussed in the light of the recent non-detection of Gl15A\,b in the CARMENES visual RV time series by \citet{trifonovetal2018}.
   
   With tens of exoplanets orbiting binary stars observed to date \citep[e.g.][]{eggenberger2010}, including a confirmed wide binary system with planetary companions orbiting both components \citep{desideraetal2014,damassoetal2015}, it is debated how the presence of stellar companions influenced the distribution of planetary orbital parameters.
   For this reason, we derived new orbital parameters for the stellar companion Gl15B, using HARPS-N RV measurements along with astrometric measurements from the WDS \citep[Washington Double Star,][]{masonetal2001} catalogue, and performed several numerical simulations to test the dynamical influence of Gl15B on the planetary system.
   
   We describe the Doppler and photometric measurements collected for the analysis in Sect. \ref{rv_time_series}, and then, in Sect. \ref{stellar_prop}, we describe the host star and binary companion updated properties. The complete analysis of the RVs and activity indices of the system is presented in Sect. \ref{spect_sec}. We then analyse the binary orbit and the perturbations produced on the two planets orbits in Sect. \ref{binary_interaction}. We summarize and discuss our findings in Sect. \ref{paperm22_conclusions}.


\section{Observations and HIRES catalog data}
\label{rv_time_series}

As part of the HADES RV programme, Gl15A has been observed from BJD $=2456166.7$  (27th August 2012) to BJD $= 2457772.4$ (18th January 2017).
The total number of data points acquired was 115, over a time span of $1605$ days. 
The HARPS-N spectra were obtained using an exposure time of $15$ minutes, and achieving an average signal-to-noise ratio ($S/N$) of $150$ at $5500$ \AA. Of the 115 epochs, $49$ were obtained within the GAPS time and $67$ within the Spanish time.
Since the simultaneous Th-Ar calibration could contaminate the Ca II H \& K lines, which are crucial in the analysis of stellar activity of M dwarfs \citep{forveilleetal2009,lovisetal2011}, the observations were gathered without it.
However, to correct for instrumental drift during the night, we used other GAPS targets spectra, gathered by the Italian team during the same nights as the Gl15A observations using the simultaneous Th-Ar calibration.

The data reduction and RV extraction were performed using the TERRA pipeline \citep[Template-Enhanced Radial velocity Re-analysis Application,][]{anglada-escudebutler2012}, which is considered to be more accurate when applied to M-dwarfs, with respect to the HARPS-N Data Reduction Software \citep[DRS,][]{lovispepe2007}. For a more thorough discussion of the DRS and TERRA performances on the HADES targets see \citet{pergeretal2017}. The rms of the TERRA RVs is $2.69$ m s$^{-1}$, while the mean internal error is $0.62$ m s $^{-1}$. The TERRA pipeline also corrected the RV data for the perspective acceleration of Gl15A, ${dv_r / t = 0.69}$ m s$^{-1}$ yr$^{-1}$.

We also use in our analysis the HIRES-Keck binned data for Gl15A, downloaded from the LCES HIRES/Keck Precision Radial Velocity Exoplanet Survey. The HIRES time series spans $6541$ days, from BJD $=2450461.8$  (13th January 1997) to BJD $= 2457002.7$ (11th December 2014). These data were newly reduced with respect of the original RVs used by \citet{howardetal2014}, and also new data have been observed with the HIRES spectrograph since the paper came out. For a full description of the catalog and data reduction see \citet{butleretal2017}. We discarded the last data point of the HIRES time series (BJD \nolinebreak$= 2457002.7$), since it is almost $\sim 10$ m s$^{-1}$ off with respect of the rest of the data.
We thus use in our analysis $169$ HIRES RVs, showing a variation of $3.26$ m s$^{-1}$ and an average internal error of $0.84$ m s$^{-1}$. The RV dataset from the Keck archive was already corrected for the perspective acceleration of the host star.

\begin{figure}
   \centering
   \includegraphics[width=9cm]{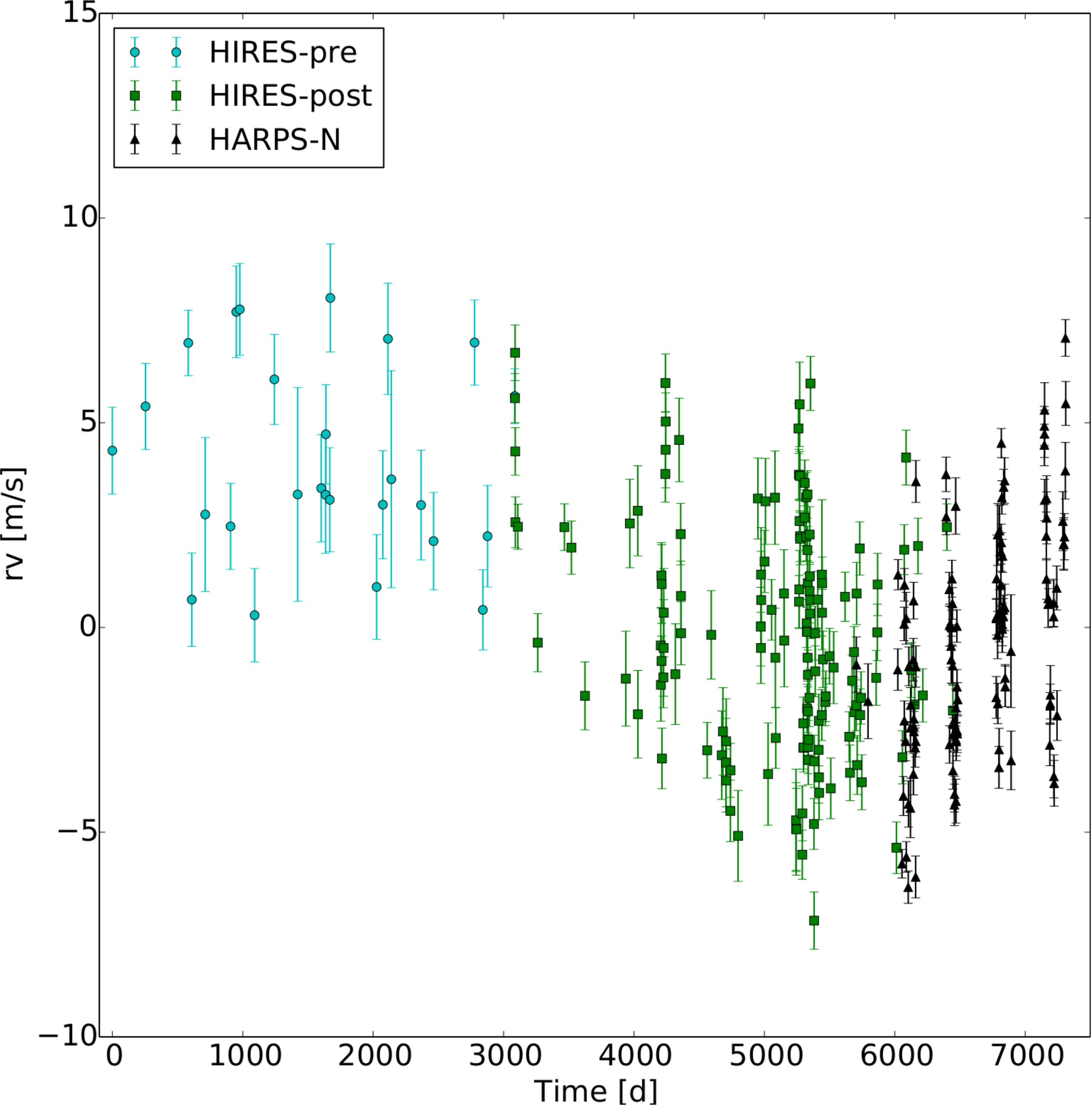}
      \caption{Combined HARPS-N and HIRES RV time series. As described in Section \ref{emcee_analysis}, we considered the HIRES data as split in two separate datasets for the purpose of the RV analysis.}
         \label{time_series}
\end{figure}

We combine the HARPS-N and HIRES RV datasets, obtaining a $290$ points time series, spanning $7310$ days. The complete RV time series has rms $=3.08$ m s$^{-1}$, mean error $=0.71$ m s $^{-1}$. Figure \ref{time_series} shows the combined RV time series, with the respective mean RV subtracted from each dataset to visually compensate for the offset expected between the measurements of the two instruments.

As a part of the analysis of the binary orbit described in Section \ref{binary_interaction}, we obtained with HARPS-N 5 RV data points of the companion star Gl15B, collected from BJD $=2457753.9$ (31st December 2016) to BJD $=2457771.9$ (18th January 2017). The rms of the time series is $1.26$ m s$^{-1}$, while the mean internal error is $2.42$ m s $^{-1}$.

As for every target in the HADES survey, Gl15A was also monitored photometrically by the EXORAP (EXOplanetary systems Robotic APT2 Photometry) program, to estimate the stellar rotation from the periodic modulation in the differential light curve.

The observations were performed by INAF-Catania Astrophysical Observatory at Serra la Nave on Mt. Etna (+14.973$^\circ$E, +37.692$^\circ$N, 1725m a.s.l.) with the APT2 telescope, which is an 80cm f/8 Ritchey-Chretien robotic telescope. The detector is a 2k$\times$2k e2v CCD 230-42, operated with standard Johnson-Cousins $UBVRI$ filters. The IDL routine \textit{aper.pro} was used to implement the aperture photometry. The data were collected from the 13th of August 2013 to the 15th of June 2017, for a total 242 and 233 data points, for the B and V photometry respectively.

\section{Stellar Properties of Gl15A and Gl15B}
\label{stellar_prop}

Gl15A is a high proper motion nearby ($\pi = 280.3 \pm 1.0$ mas) early-M dwarf, of type M1. We used the stellar parameters published by \citet{maldonadoetal2017}, which were calculated applying the empirical relations by \citet{maldonadoetal2015} on the same HARPS-N spectra from which we derived the RV time series.
This technique calculates stellar temperatures from ratios of pseudo-equivalent widths of spectral features, and calibrate the metallicity on combinations and ratios of different features. Although such techniques are mainly used for solar-type stars, \citet{maldonadoetal2015} proved them to be just as effective on low-mass stars.

\citet{howardetal2014} derived a rotational period of $43.82 \pm 0.56$ d, both from their Keck-HIRES measurements of the $S_\text{HK}$ index and their APT photometric observations at Fairborn Observatory. Recently \citet{suarezmascarenoetal2017b} analysed the potential signatures of magnetic activity in the CaII H$\&$K and H$\alpha$ activity indicators of the HADES M-dwarfs sample. For Gl15A they computed a mean level of chromospheric emission $\log R'_\text{HK} = -5.27 \pm 0.04$, and a rotation period $P_\text{rot} = 45.0 \pm 4.4$ d, fully consistent with the value found by \citet{howardetal2014}. The definition of $\log R'_\text{HK}$ can be extended for application on M-dwarfs spectra \citep[and references therein]{suarezmascarenoetal2017b}.

All the stellar parameters for Gl15A used in this work are summarized in the left column of Table \ref{star_par}. We can see that most of them are fully consistent with the values used by \citet{howardetal2014}.

\begin{table}
\caption{Stellar parameters for the two stars Gl15A and Gl15B}             
\small
\label{star_par}      
\centering                          
\begin{tabular}{lcc}        
\hline\hline                 
Parameter & Gl15A & Gl15B\\    
\hline                        
   Spectral Type & M1 \tablefootmark{a} & M3.5 \tablefootmark{g} \\      
   $T_{\text{eff}}$ $[$K$]$ & $3607 \pm 68$ \tablefootmark{a} & $3304 \pm 70$ \tablefootmark{g} \\
   $[$Fe$/$H$]$ $[$dex$]$ & $-0.34 \pm 0.09$ \tablefootmark{a} & $-0.37 \pm 0.10$ \tablefootmark{g} \\
   Mass $[$M$_\odot]$ & $0.38 \pm 0.05$ \tablefootmark{a} & $0.15 \pm 0.02$ \tablefootmark{g} \\
   Radius $[$R$_\odot]$ & $0.38 \pm 0.05$ \tablefootmark{a} & $0.18 \pm 0.03$ \tablefootmark{g} \\
   $\log g$ $[$cgs$]$ & $4.87 \pm 0.04$ \tablefootmark{a} & $5.08 \pm 0.15$ \tablefootmark{g} \\
   $\log \text{L}_*/\text{L}_\odot$ & $-1.655 \pm 0.112$ \tablefootmark{a} & $  -3.070 \pm 1.221 $ \tablefootmark{g} \\
   $v \sin i$ $[$km s$^{-1}]$ & $1.09 \pm 0.79$ \tablefootmark{a} \\
   $\log R'_\text{HK}$ & $-5.27 \pm 0.04$ \tablefootmark{b} \\
\hline                                   
$\alpha$ (J2000) & 00$^h$:18$^m$:20.5$^s$ \tablefootmark{c} \\
$\delta$ (J2000) & +44$^\circ$:01$'$:19$''$ \tablefootmark{c} \\
$B-V$ $[\text{mag}]$ & 1.55 \tablefootmark{d}\\
$V$ $[\text{mag}]$ & 8.13 \tablefootmark{e} \\
$J$ $[\text{mag}]$ & 5.25 \tablefootmark{f} \\
$H$ $[\text{mag}]$ & 4.48 \tablefootmark{f} \\
$K$ $[\text{mag}]$ & 4.02 \tablefootmark{f} \\
$\pi$ $[\text{arcsec}]$ & $0.2803 \pm 0.0010$ \tablefootmark{c} \\
$\mu_\alpha$ $[\text{arcsec yr}^{-1}]$ & 2.888 \tablefootmark{c} \\
$\mu_\delta$ $[\text{arcsec yr}^{-1}]$ & 0.409 \tablefootmark{c} \\
\hline
\end{tabular}
\tablefoot{\tablefoottext{a}{\citet{maldonadoetal2017}}; \tablefoottext{b}{\citet{suarezmascarenoetal2017b}};  \tablefoottext{c}{\citet{esa1997}}; \tablefoottext{d}{\citet{leggett1992}}; \tablefoottext{e}{\citet{hogetal2000}}; \tablefoottext{f}{\citet{cutrietal2003}}; \tablefoottext{g}{This work}}
\end{table}

The stellar companion of Gl15A, Gl15B, is a type M3.5 dwarf whose orbit was measured astrometrically by \citet{lippincott1972}. For the purpose of our orbital analysis in Sect. \ref{binary_interaction}, we took 5 HARPS-N spectra of Gl15B during the last observing season at TNG (see next Section) and we applied on them the \citet{maldonadoetal2015}'s techniques to calculate updated stellar properties. The derived values are listed in the right column of Tab. \ref{star_par}.

\subsection{Photometric analysis}

In order to identify the potential modulation in the stellar photometry due to the presence of active regions, we analyse with the GLS periodogram \citep{zechkur2009} the B and V time series collected within the EXORAP survey, composed of 242 and 233 points respectively, taken over five seasons from the 13th of August 2013 to the 15th of June 2017. No evident periodicity is found in either time series, in contrast with the findings of \citet{howardetal2014} who found a clear signal at 43.82 days (see their Figure 4), identified as the rotation period of the star,  with a corresponding signal seen in the CaII activity index. The discrepancy between the two analysis may be due to the different photometric precision of the observations: the amplitude of the signal found by \citet{howardetal2014} was only of $6$ mmag, which is below the sensitivity of the EXORAP survey for targets in the magnitude range of Gl15A.

\section{Spectroscopic data analysis}
\label{spect_sec}

In Figure \ref{time_series} an evident long period signal can be seen. \citet{howardetal2014} identified in their HIRES data a hint of a long period decreasing trend, which they included in their model as a constant negative acceleration of $-0.26 \pm 0.09$ m s$^{-1}$ yr$^{-1}$, and concluded it to be consistent with the orbit of the Gl15AB system as calculated by \citet{lippincott1972}. But considering also the new HARPS-N data, it becomes clear how the long term RV variation cannot be modeled as a linear trend. 

Even if it could no longer be treated as a constant acceleration, the long period signal could still be due to the binary reflex motion of Gl15A, so we investigate this possibility. 
To model the possible RV signal due to the stellar companion we follow the procedure used by \citet{kippingetal2011} to study the presence of a long-period companion in the HAT-P-31 system: they modeled the long-period signal as a quadratic trend, and then derived a range of orbital parameters from the quadratic coefficients. The ratio between the semi-amplitude, $K_B$, and period, $P_B$, of the companion signal is given by the second-order term of the trend, $\ddot \gamma$, as (their Equation (4)):
\begin{equation}
{K_B \over P_B^2} = {\ddot \gamma \over 4 \pi^2},
\end{equation}
and, since $K_B$ depends on the orbital period and on the mass of the two stars, the orbital period can be derived as a function of the second order term and the masses $P_B = P_B(\ddot \gamma, M_A, M_B)$.

From the fit of the complete RV time series we obtain: $\ddot \gamma = 3.95 \times 10^{-7}$ m s$^{-1}$ day$^{-2}$. Using the stellar masses listed in Table \ref{star_par}, the resulting period is $P_B \simeq 680$ yr, which would correspond to a semi-major axis $a_B \simeq 63$ AU. This solution is clearly unphysical, since the presumed semi-major axis is less than half the observed orbital separation of the two objects.

Moreover, several high-contrast imaging surveys ruled out the presence of additional co-moving stellar objects close to Gl15A: \citet{vanburenetal1998} excluded the possibility of objects with $M_\star > 0.084$ M$_\odot$ at separations of $9-36$ AU, while \citet{tanneretal2010} ruled out objects up to a magnitude contrast of $\Delta K_s \simeq 6.95$ mag within $1''$ (3.57 AU) and $\Delta K_s \simeq 10.24$ mag within $5''$ (17.8 AU).

The long-period signal observed in the RV time series is therefore very unlikely to be caused either by Gl15B or by additional stellar companions. Instead a long-period planetary-mass companion orbiting around Gl15A\,could be a possible explanation of the observed signal. We now investigate this hypothesis, by analysing both the potential presence of a Keplerian signal in the RVs and the stellar activity signals in the chromospheric indicators, which could also cause long-period variations due to the star magnetic cycle.


\subsection{The MCMC model}

Bearing in mind that a signal tightly linked to the stellar rotation period is clearly present in the RV data, as shown by \citet{howardetal2014}, we have selected the Gaussian process (GP) regression as a useful, and physically robust, tool both to investigate the presence of periodicities in the chromospheric activity indicators and to mitigate the stellar activity contribution to the RV variability. The GP regression is becoming a commonly used method to suppress the stellar activity correlated "noise" in RV timeseries \citep[e.g.][and references therein]{dumusque2017}, especially when adopting a \textit{quasi-periodic} covariance function. This function is described by four parameters, called \textit{hyperparameters}, and it can model some of the physical phenomena underlying the stellar noise through a simple, but efficient, analytical representation. 
The quasi-periodic kernel is described by the covariance matrix
\begin{eqnarray}
\label{eq_gpker}
K(t, t^{\prime}) = h^2\cdot\exp\bigg[-\frac{(t-t^{\prime})^2}{2\lambda^2} - \frac{sin^{2}(\dfrac{\pi(t-t^{\prime})}{\theta})}{2w^2}\bigg] + \nonumber \\
+\, (\sigma^{2}_{\rm instr, data}(t)\,+\,\sigma_{\rm instr,jit}^{2})\cdot\delta_{\rm t, t^{\prime}},
\end{eqnarray}
where $t$ and $t^{\prime}$ indicate two different epochs.

This kernel is composed by a periodic and an exponential decay term, so that it can model a recurrent signal linked to stellar rotation, also taking into account the finite-lifetime of the active regions. Such approach is therefore particularly suitable when modeling signals of short-term timescales, as those modulated by the stellar rotation period.

About the covariance function hyperparameters, $h$ is the amplitude of the correlations; $\theta$ represents the rotation period of the star; $w$ is the length scale of the periodic component, linked to the size evolution of the active regions; and $\lambda$ is the correlation decay timescale, that can be related to the active regions lifetime.
In Eq. \ref{eq_gpker}, $\sigma_{\rm instr, data}(t)$ is the data internal error at time $t$ for each instrument; $\sigma_{\rm instr, jit}$ is the additional uncorrelated 'jitter' term, one for each instrument, that we add in quadrature to the internal errors in the analysis of the RV datasets to take into account additional instrumental effects and noise sources neither included in $\sigma_{\rm instr, data}(t)$ nor modeled by the quasi-periodic kernel; $\delta_{\rm t, t^{\prime}}$ is the Kronecker delta function.

In the GP framework, the log-likelihood function to be maximized by the Markov chain Monte Carlo (MCMC) procedure is:
\begin{equation}
\label{eq_loglik}
\ln \mathcal{L} = -\frac{n}{2}\ln(2\pi) - \frac{1}{2}\ln(det\,\mathbf{K}) - \frac{1}{2}\overline{r}^{T}\cdot\mathbf{K}^{-1}\cdot\overline{r},
\end{equation}
where $n$ is the number of the data points, \textbf{K} is the covariance matrix built from the covariance function in Equation (\ref{eq_gpker}), and $\overline{r}$ is the data vector.

We use the publicly available \texttt{emcee} algorithm \citep{foreman13} to perform the MCMC analysis, and the publicly available \texttt{GEORGE} Python library to perform the GP fitting within the MCMC framework \citep{ambi14}. We used 150 random walkers to sample the parameter space. The posterior distributions have been derived after applying a burn-in as explained in \citet{eastman13} (and references therein). To evaluate the convergence of the different MCMC analyses we calculate the integrated correlation time for each of the parameters, and run the code a number of steps around 100-1000 times the autocorrelation times of all the parameters, depending on the complexity of each analysis. This ensures that the \texttt{emcee} walkers thoroughly sample the parameter space \citep{foreman13}.

\subsection{Analysis of the activity indexes}
\label{gp_act_ind}

We first investigate both the HIRES and HARPS-N CaII H$\&$K and H$\alpha$ index time series, in order to test the potential stellar origin of the long period variation seen in the combined RV times series shown in Figure \ref{time_series}. No long period trend is found in either the HIRES or HARPS-N datasets. \citet{suarezmascarenoetal2017b} found a magnetic-cycle type periodicity at $P_\text{cycle} = 2.8 \pm 0.5$ yr in the HARPS-N CaII H$\&$K and H$\alpha$ indexes, even if we don't find any similar signal in the respective HIRES time series. Nevertheless the period of this cycle is far from the time span of the long-period signal we care to investigate, so it could not be the origin of it.
Another clue of the stellar origin of the long-period modulation of the RVs could be the correlation between the activity indexes and RV time series, which we computed via the Spearman’s rank correlation coefficients. No significant correlation was identified ($\left | \rho \right | < 0.5$ for all the indexes).

These suggest that the long-period signal is not due to the star magnetic cycle or chromospheric activity, and reinforces the hypothesis of it be due to a wide-orbit planetary companion.

\begin{figure}
   \centering
   \includegraphics[width=9cm]{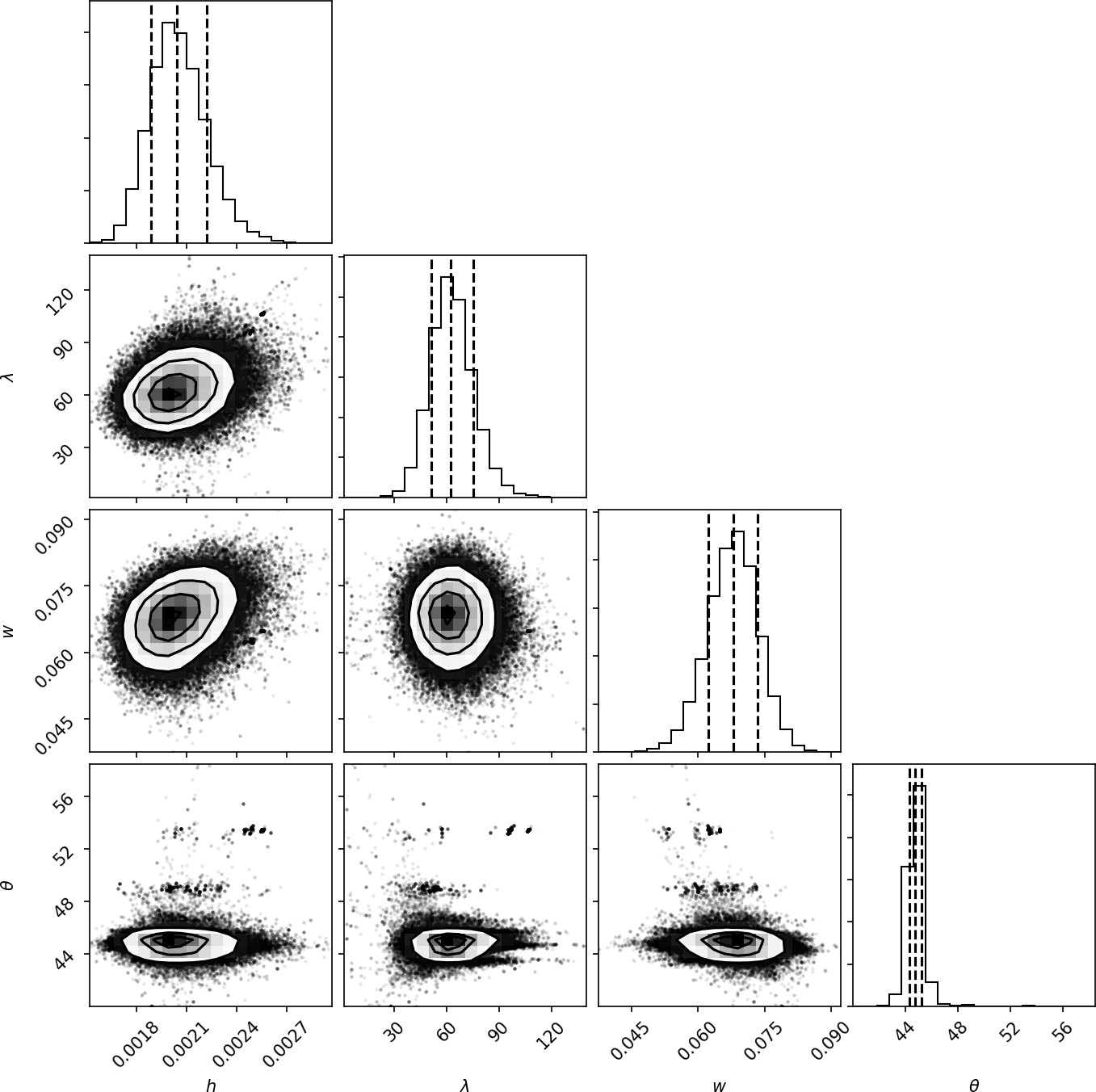} \\
   \includegraphics[width=9cm]{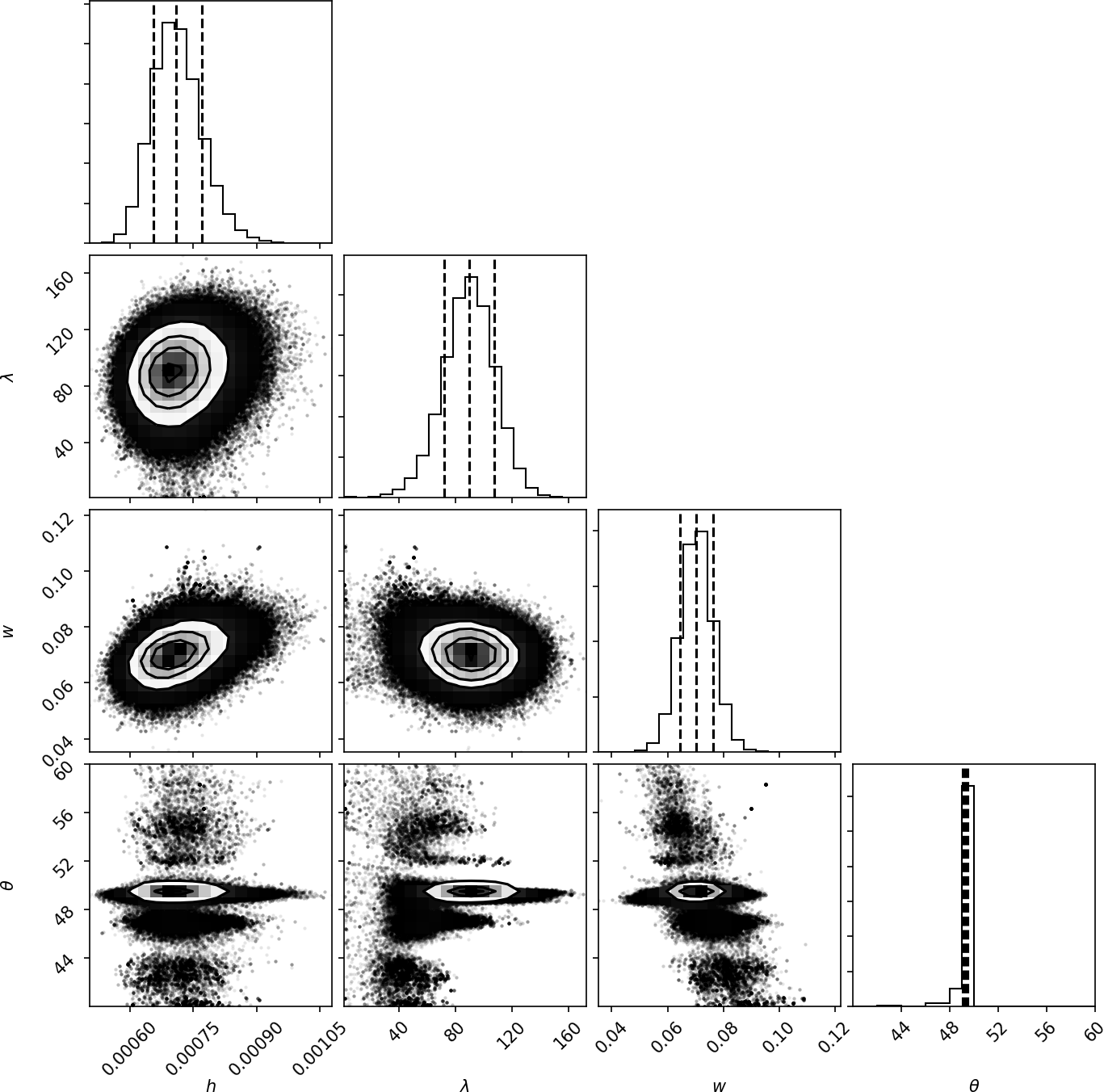}
      \caption{Posterior distributions of the fitted (hyper-)parameters of the GP quasi-periodic model applied to the timeseries of CaII H$\&$K (upper panel) and H$\alpha$ (lower panel) activity indexes. The vertical lines denote the median (solid) and the 16$^{th}-84^{th}$ percentiles (dashed).}
         \label{fig_3-1}
\end{figure}

To further test the effect of the activity of Gl15A, we then investigate the stellar activity behaviour over the four seasons covered by HARPS-N observations by analysing chromospheric activity indicators based on the Ca II H and K, and H$\alpha$ spectroscopic lines. They were extracted using the definition of the line cores and of the reference intervals given in \cite{gomesdasilva11}. We analyse here only measurements obtained from HARPS-N spectra because they represent an homogeneous dataset. By spanning more than 1600 days, the HARPS-N data alone can provide robust insights into the long- and short-term stellar activity variability. 
The average S/N was 18 for the CaII H\&K index and $213$ for the H$\alpha$.

We performed an analysis of the activity indicators based on a Gaussian process (GP) regression, as detailed in the previous Section. By adopting the same covariance function (Eq. \ref{eq_gpker}) we use to model the stellar contribution present in the RV variations in the following section, our primary goal is to investigate some properties of the stellar activity and use them to constrain some parameter priors in the analysis of the radial velocities. This represents a reasonable expectation, because neither the activity indicators nor the RVs have been pre-whitened, and the variability patterns in the former could be present with similar properties in the latter, as was noticed by \citet{afferetal2016} during the analysis of another HADES target. In this sense, results from the analysis of the activity indicators can be used to train the GP regression of the RVs, by keeping unchanged the way the stellar activity contribution is modeled. 
Here we use Eq. \ref{eq_gpker} to describe the variability correlated with the stellar rotation period $P_{\rm rot}$, by adopting a uniform prior for the corresponding hyperparameter $\theta$ which is constrained between 40 and 60 days (the list of priors is shown in Table \ref{tab-gp-act-ind}).

The MCMC analysis was stopped after running for around 1000 times the highest autocorrelation time. The posterior distributions of the model parameters are shown in Fig. \ref{fig_3-1}, and the best-fit estimates are listed in Table \nolinebreak \ref{tab-gp-act-ind}. We note that the estimates of the stellar rotation periods are well constrained but slightly different for the two activity indicators. The value found from the CaII H$\&$K index is very similar to that found by \cite{howardetal2014} for the CaII H$\&$K S-index derived from HIRES spectra. The hyper-parameter $\theta$ found in the analyses of the two time series is in very good agreement with the rotation period found by \citet{suarezmascarenoetal2017b} with a GLS analysis of the activity indexes time series of Gl15A.

\begin{table}
   \caption[]{Priors and best-fit results for the Gaussian process regression analysis of the chromospheric activity indicators extracted from the HARPS-N spectra of Gl15A.}
          \label{tab-gp-act-ind}
          \centering
         \small
    \begin{tabular}{l l l l}
             \hline
             \noalign{\smallskip}
             Jump parameter     &  prior & \multicolumn{2}{c}{Best-fit value}  \\
             \noalign{\smallskip}
                 &   & CaII H$\&$K & H$\alpha$  \\
             \hline
             \noalign{\smallskip}
             $h$  & $\mathcal{U}$(0,0.5) & 0.0020$\pm0.0002$ & $(7.1\pm0.6) \times 10^{-4}$\\
             \noalign{\smallskip}
             $\lambda$ [days] & $\mathcal{U}$(0,10\,000) & 62$^{+13}_{-11}$ & 90$\pm18$\\
             \noalign{\smallskip}
             $w$ & $\mathcal{U}$(0,1) & 0.068$^{+0.005}_{-0.006}$ & 0.070$\pm0.006$ \\
             \noalign{\smallskip}
             $\theta$ [days] & $\mathcal{U}(40,60)$ & 44.7$^{+0.5}_{-0.4}$ & 49.3$\pm0.2$ \\
             \noalign{\smallskip}
             \hline
      \end{tabular}
\end{table}

\subsection{Analysis of the combined HIRES and HARPS-N RV time series}
\label{emcee_analysis}

\begin{figure}
\centering
\includegraphics[width=1.0\hsize]{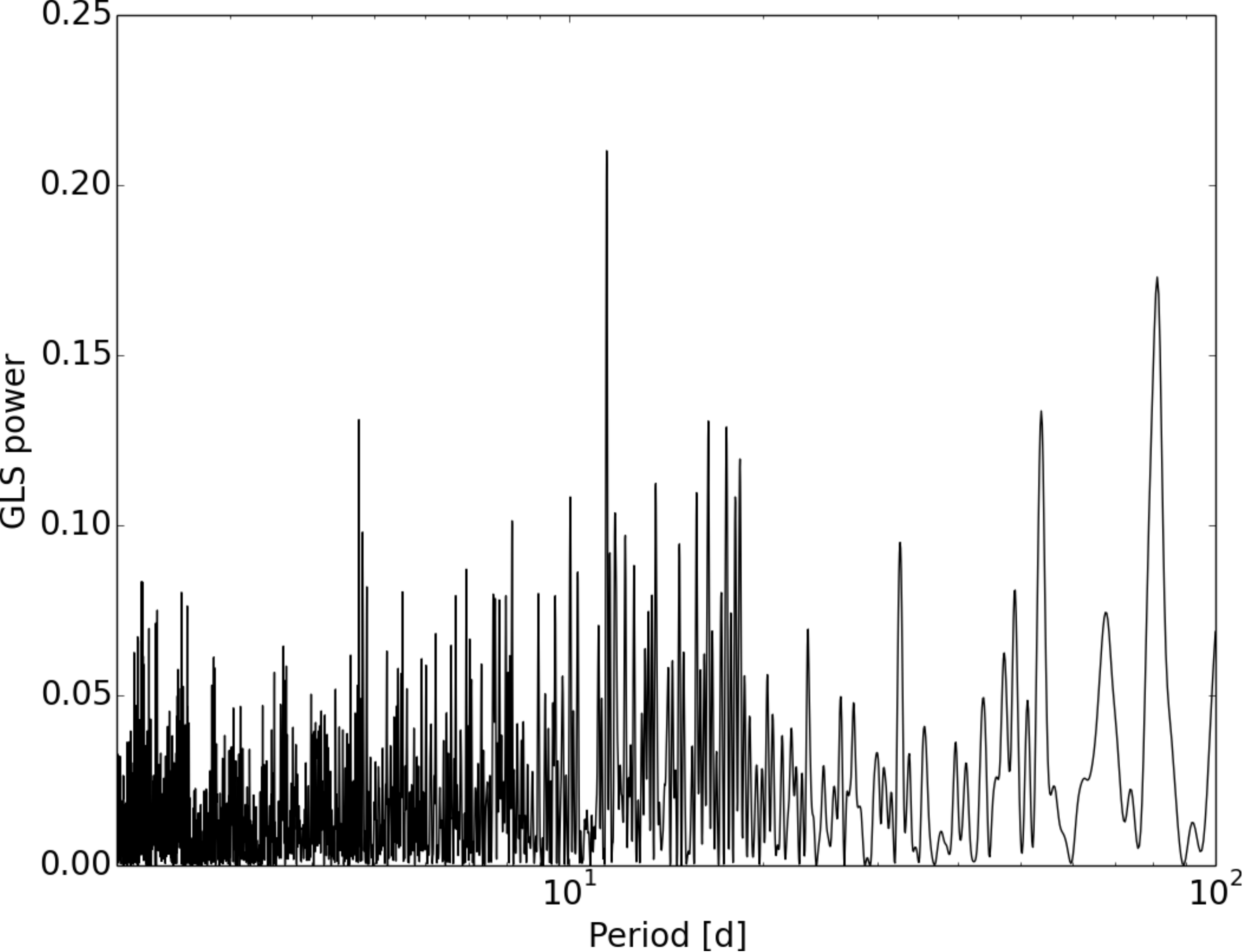}
\caption{GLS periodogram of the complete HARPS-N RV dataset}
\label{fig_periodogram}
\end{figure}

\begin{figure}
\centering
\subfloat[][]
 {\includegraphics[width=1.0\hsize]{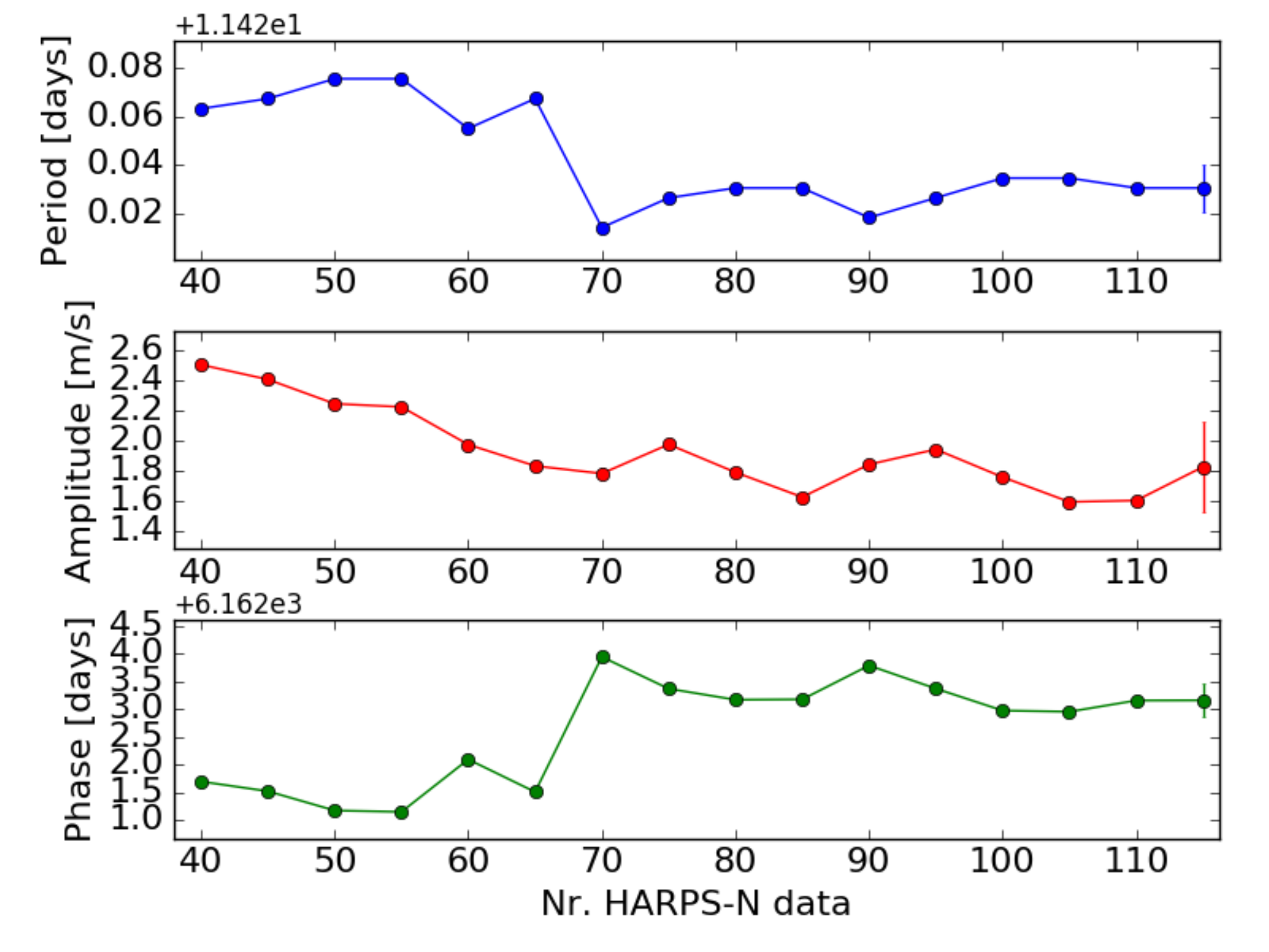}}\\
\subfloat[][]
 {\includegraphics[width=1.0\hsize]{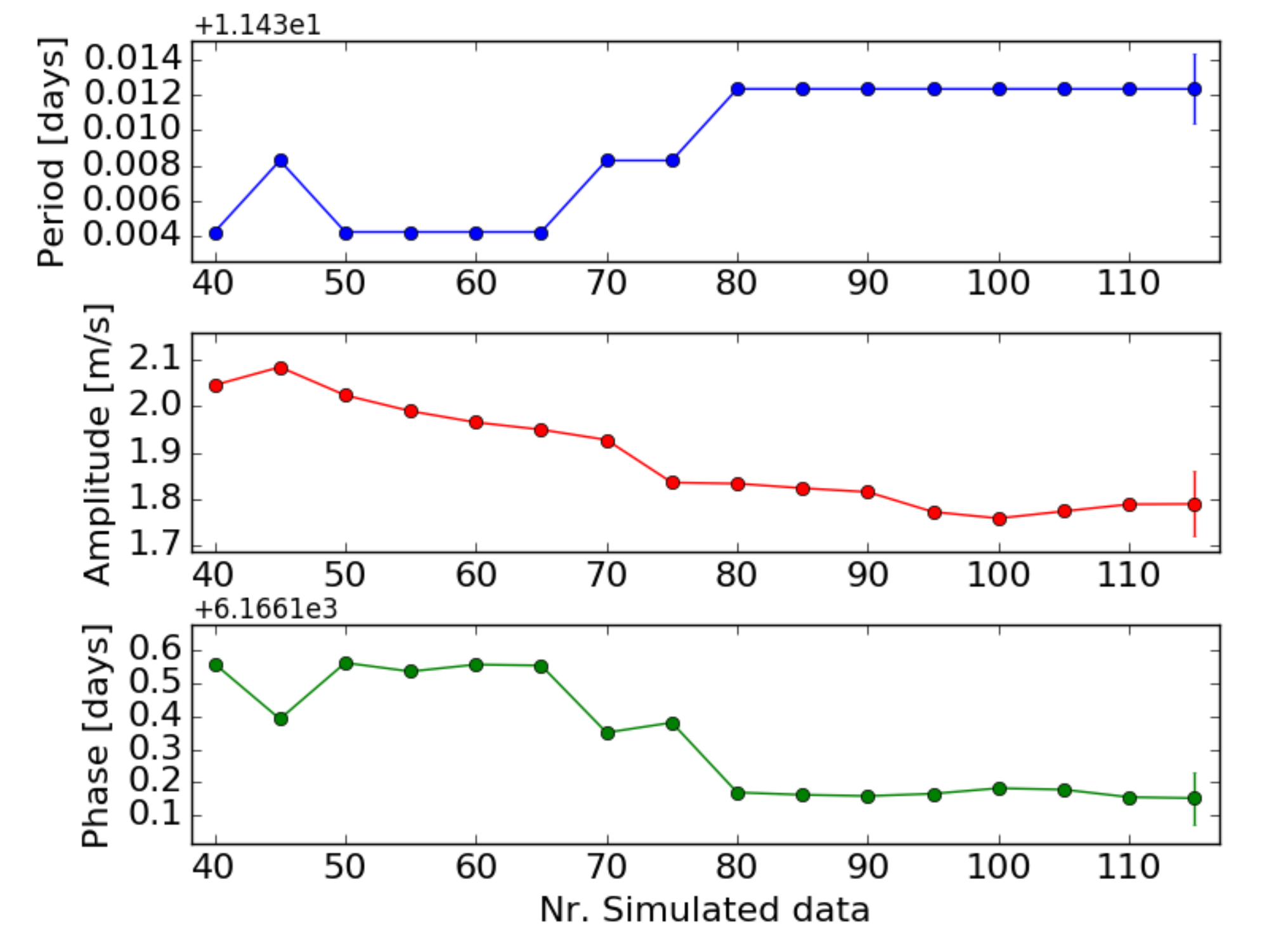}}
\caption{a) Evolution of the GLS orbital parameters as a function of the number of HARPS-N RV measurements; b) evolution of the GLS orbital parameters as a function of the number of simulated data. The error bars on the final points are shown as references.}
\label{fig_par_evolution}
\end{figure}

First we study the HARPS-N RV time series for confirmation of the presence of the Gl15A\,b signal found by \citet{howardetal2014}. We perform a GLS analysis to identify any periodicity in our data in $P \in [2,100]$ d. In the periodogram, shown in Fig. \ref{fig_periodogram}, we can see that the $P = 11.44$ d period of planet Gl15A\,b is clearly recovered. As a confirmation of the coherence of the signal throughout the entire HARPS-N time series, Fig. \ref{fig_par_evolution}a shows the evolution of the period, amplitude and phase recovered by the GLS periodogram as a function of the number of observations: we can clearly see how from $\sim70$ forward the period and phase of the signal remain stable, with small oscillations of the order of the final error on the parameters, even if the remainder of the observations covers almost one and a half years. We also studied the periodogram of the CaII H\&K and H$\alpha$ HARPS-N time series, and found no significant signal at periods close to the inner planet period $P_b$, thus confirming its planetary nature as announced by \citet{howardetal2014}.

It is worth noticing that, as can be seen in Fig. \ref{fig_periodogram}, the final amplitude recovered by the GLS periodogram on the HARPS-N dataset, $K_\text{b} = 1.82 \pm 0.31$ m s$^{-1}$, is smaller than the one published by \citet{howardetal2014}, $K_\text{b,How} = 2.94 \pm 0.28$ m s$^{-1}$. The decreasing behaviour of the amplitude recovered by GLS with increasing number of observation is not unexpected, as the sampling of the signal strongly influence the periodogram structure and fit, and we tested that a similar behaviour can be observed  also in simulated datasets with the same epochs as our HARPS-N time series but containing only the planetary signal and white noise, as shown in \ref{fig_par_evolution}b. The lower amplitude value than that found by \citet{howardetal2014} can also be explained similarly, since the HARPS-N time series is composed of five years of continuous high-cadence observations of the target, while the dataset used by \citet{howardetal2014} consisted in mostly sparse measurements with an intensive high-cadence monitoring only in the last seasons, which could affect the signal recovery. Moreover it is worth noticing that the HIRES data from the now public LCES HIRES/Keck archive have been reduced with a different and more effective technique \citep{butleretal2017} than the data used by \citet{howardetal2014}, and performing a GLS analysis of the two time series we obtain the same peak periodicity but an amplitude $K_b$ $\sim 23 \%$ smaller in the new archive data.

Since no other short period signal emerges from the GLS analysis of the RVs, we focus our attention on the study of the long-period signal, which we assume to be due to a planetary-mass wide-orbit companion.

To estimate the orbital and physical parameters of the known planet Gl15A\,b and the candidate companion, hereafter Gl15A\,c, we have performed a Markov chain Monte Carlo analysis of the RVs.
Following the prescription of \cite{howardetal2014}, we model the RVs dividing the HIRES dataset in a pre-upgrade and a post-upgrade sublist, due to the HIRES CCD upgrade occurred on August 2004. Each subsample is then treated as an independent dataset, with its own zero-point offset and additional jitter term.
In this case, the $\overline{r}$ in Eq. \ref{eq_loglik} represents the RV residuals, obtained by subtracting the Keplerian signal(s) from the original RV dataset.

The general form for the models that we tested in this work is given by the equation
\begin{eqnarray}\label{eq:4-3}
\Delta RV(t) =  \gamma_{\rm instr} + \sum_{\rm j=1}^{n_{\rm planet}} \Delta RV_{\rm Kep,j}(t)
                   +\, \dot{\Delta RV(t)} +
\nonumber \\
			+\, \Delta RV(t)_{\rm(activity,\, short-term)} =
\nonumber \\
= \gamma_{\rm instr.} \hspace{-.05cm} + \hspace{-.05cm} \sum_{\rm j=1}^{n_{\rm planet}}\hspace{-.05cm} K_{\rm j}\large[\cos(\nu_{j}(t, e_{\rm j}, T_{\rm 0,j}, \textit{P}_{\rm j}) + \omega_{\rm j}) \hspace{-.05cm} +  e_{\rm j}\cos(\omega_{\rm j})\large]  \hspace{-.05cm} + 
\nonumber \\
              + \dot{\Delta RV(t)} + \hspace{-.05cm} GP
\end{eqnarray}
where $n_{\rm planet}=1,2$; $\nu$ is a function of time \textit{t}, time of the inferior conjuntion $T_{\rm 0,j}$, orbital period $\textit{P}_{\rm j}$, eccentricity \textit{e} and argument of periastron $\omega_{\rm j}$; $\gamma_{\rm instr}$ is the RV offset, one for each instrument; $GP$ is the stellar noise modeled with the Gaussian Process. The term $\dot{\Delta RV(t)}$ is the residual acceleration of the system, which can include also the RV acceleration due to the orbital motion within the Gl15AB binary system, which is difficult to predict due to the large uncertainties in the orbital parameters (as we will discuss in Sec. \ref{orbit_mod}).
Instead of fitting separately $e_{\rm j}$ and $\omega_{\rm j}$, we use the auxiliary parameters
$C_{\rm j} = \sqrt{e_{\rm j}}\cdot \cos \omega_{\rm j}$ and $S_{\rm j} = \sqrt{e_{\rm j}} \cdot \sin \omega_{\rm j}$ to reduce the covariance between $e_{\rm j}$ and $\omega_{\rm j}$.
\citet{howardetal2014} found the eccentricity of planet Gl15A\,b to be consistent with zero, and concluded a circular orbit to be the best fit for the data. To confirm this conclusion or to evaluate the real value of the eccentricity $e_b$, we assume in our analysis a Keplerian orbit for both planets. The eventual eccentricity orbit for Gl15A\,b would also be of interest as a potential consequence of the dynamical interaction of planet $b$ with Gl15B, as we will discuss in Sec. \ref{binary_interaction}.

\begin{figure*}
   \centering
   \includegraphics[width=16cm]{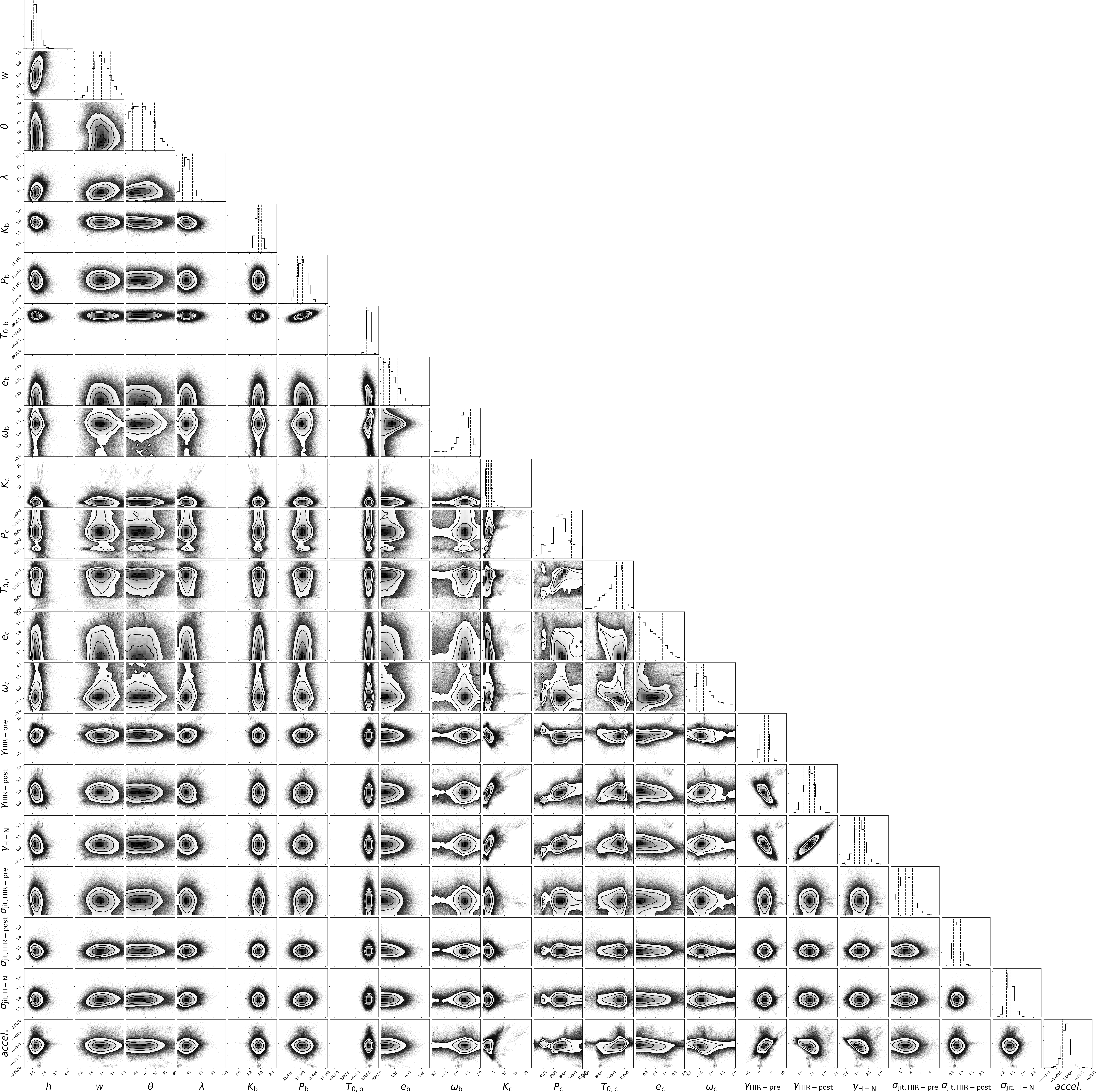}
      \caption{Posterior distributions of the fitted (hyper-)parameters of the two-planet model, where the stellar correlated noise has been modeled with a GP regression using a quasi-periodic kernel. On the \textit{y}-axis is shown the logarithm of the product between the likelihood and the prior. The vertical lines denote the median (solid) and the 16$^{th}-84^{th}$ percentiles (dashed).}
         \label{fig_postdist}
\end{figure*}

Except for very few parameters, for our analysis we assumed uniform priors.
Our choice for the range of $\lambda$ is justified by the results obtained from the GP analysis of the Ca\,II and H$\alpha$ spectroscopic activity indexes (see Table \ref{tab-gp-act-ind}), and from a preliminary, quick MCMC test on the data, which showed that the chains were converging towards values not far from the expected stellar rotation period.  
For the semi-amplitudes $K$ of the Doppler signals we used the modified invariant scale prior:
\begin{equation}
f(K) = \frac{1}{K+K_{0}} \frac{1}{\ln(1+K_{max}/K_{0})}, 
\end{equation}
following the prescription in \cite{gregory10}.
For the orbital period of Gl15A\,b there was a strong constraint from the results of \citet{howardetal2014}, confirmed by our GLS analysis previously discussed. Nevertheless, since the same dataset from \citet{howardetal2014} is used in our analysis, it would be improper to use this results as an informative prior. We instead adopted an uninformative prior over a small period interval centered on the value found by \citet{howardetal2014}, e.g. $P_b \in [10.9, 12.0]$.
For the orbital period of the candidate planet Gl15A\,c we used a logarithmic prior to assume all the orders of magnitude equally likely.

The MCMC analysis was stopped after running for around 300 times the highest autocorrelation time. Table \ref{tab_planpar} summarizes the best-fit values for each of the 19 free parameters of our model, together with the details of the prior distributions used to draw the samples. The posterior distributions of the model parameters are shown in Fig. \ref{fig_postdist}.

About the best-fit values for the free parameters, we first note that the semi-amplitude of the Doppler signal induced by Gl15A\,b is even lower than the estimate obtained from the GLS periodogram, differing from the estimate of \cite{howardetal2014} (2.94$\pm$0.28 \ms) for more than 3.5-$\sigma$, resulting in a lower minimum mass. In addition to the previously discussed effect, this is a consequence of our global model, where planetary signals are fitted jointly with a term describing the stellar correlated noise, and also explains the lower values of $\sigma_{\rm HIRES, jit}$ required by our fit. 

Our dataset does not cover a complete orbit of the outer planet Gl15A\,c, then we cannot reliably constrain the eccentricity, which appears to be significant within less than 1.5-$\sigma$ ($e\,<0.40$ at a 68$\%$ level of confidence). Our estimate for the minimum mass of Gl15A\,c places this companion in the group of the so-called super-Neptunes, as are generally referred planets in the $[20,80]$ M$_\oplus$ mass range. 

The eccentricity of Gl15A\,b also resulted consistent with zero within 1.5-$\sigma$, with a much lower value ($e\,<0.13$ at a 68$\%$ level of confidence). Our analysis thus agrees with the results by \citet{howardetal2014}, who concluded that there was no evidence for an eccentric orbit to be preferable to a circular one.

\begin{figure}
   \centering
   \subfloat[][]
   {\includegraphics[width=0.9\hsize]{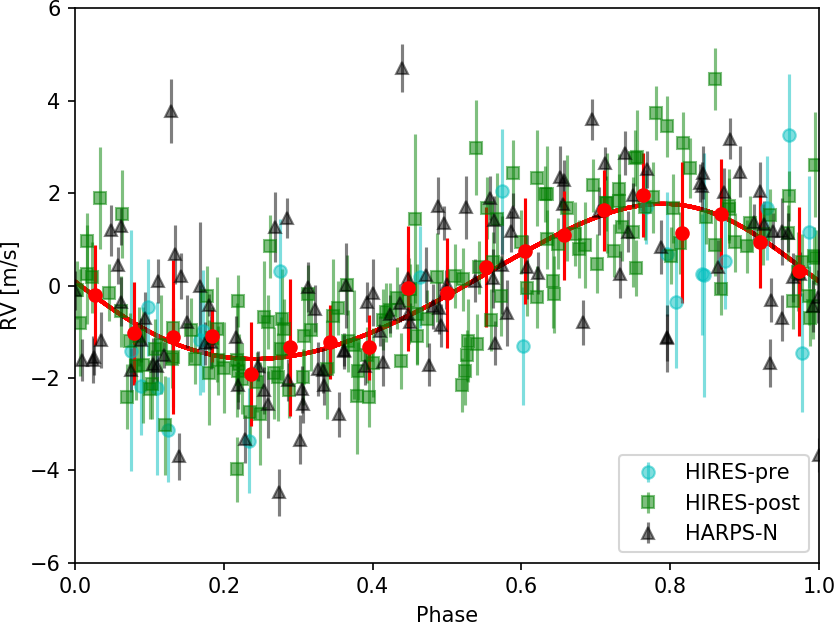}} \\
   \subfloat[][]
   {\includegraphics[width=0.9\hsize]{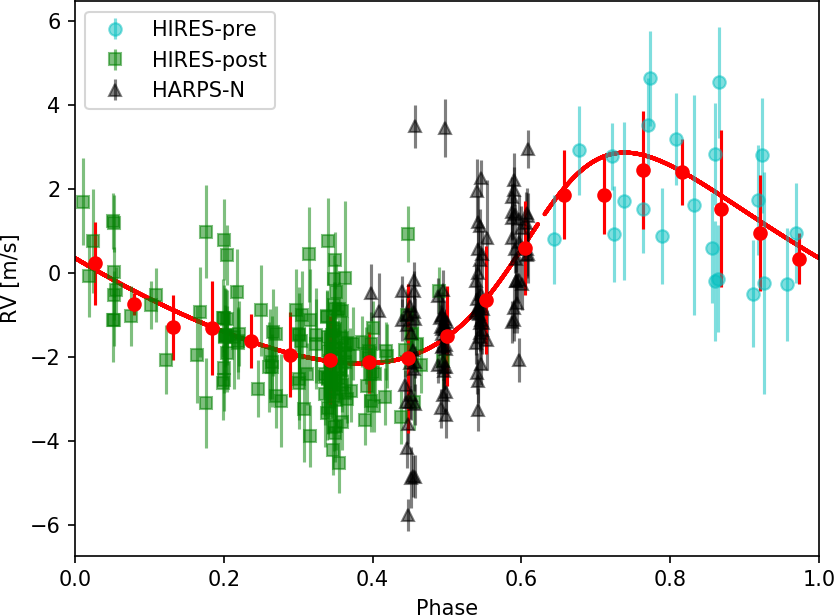}}
      \caption{Phase folded  RV curves for a) Gl15A\,b and b) Gl15A\,c. Each curve shows the residuals after the subtraction of the other planet and the stellar correlated  signal. The red curve represents the best-fit Keplerian orbit, while the red dots and error bars represent the binned averages and standard deviations of the RVs.}
         \label{fig_phasefold}
\end{figure}

\begin{figure}
   \centering
   \includegraphics[width=0.9\hsize]{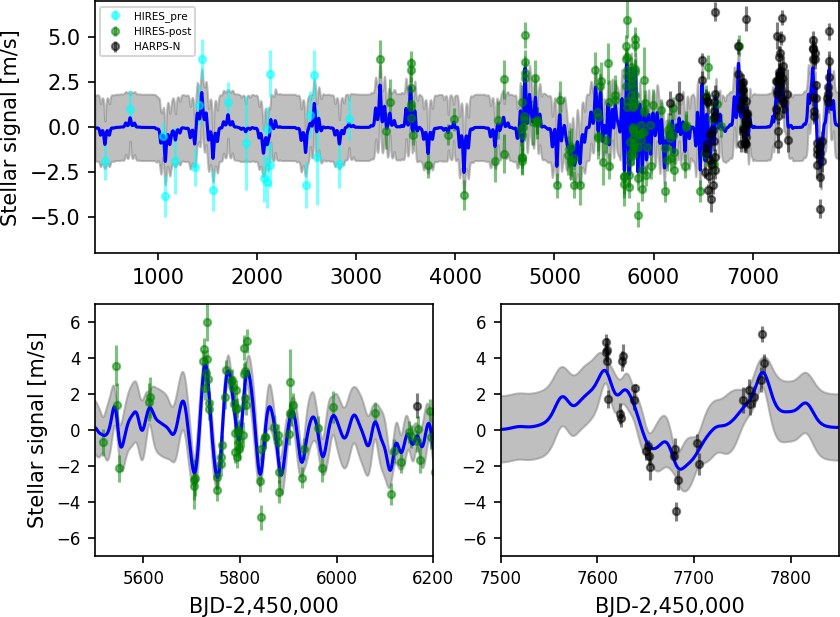}
      \caption{Upper panel: best fit stellar quasi-periodic signal (blue line) compared to the RV residuals. Lower panels: blow-up of the high-cadence HIRES/KECK observations (left) and of the last HARPS-N observing season (right).}
         \label{fig_starsign}
\end{figure}

Fig. \ref{fig_phasefold} shows the RV curves folded at the best-fit orbital periods for the known planet Gl15A\,b and the detected outer companion Gl15A\,c. In Fig. \ref{fig_starsign} we show the RV residuals, after the two Keplerians have been removed, with superposed the best-fit stellar correlated, quasi-periodic signal.

As for the GP hyper-parameters, the stellar rotation period $\theta$ assumes a value which is in agreement with that expected, and the higher uncertainties than those for the activity indicators (Tab. \ref{tab-gp-act-ind}) should be due to the longer timespan covered by the RVs, which include HIRES data. The longer timespan should also explain why the active-region evolutionary timescale $\lambda$ sets on a value less than those found for the activity indicators. Without any ancillary data available as photometry, covering the same timespan of the RVs, we cannot conclude if the value for $\lambda$ is actually physically robust, but a value not far from the rotation period seems not unreliable, as it was observed before in previous studies of different systems \citep[e.g.][]{afferetal2016}. Also, the $h$ hyper-parameter, corresponding to the mean amplitude of the stellar signal, is fully compatible with the expected value derived by \citet{suarezmascarenoetal2017b} from the mean activity level of the star, $\log R'_\text{HK}$, that is $K_\text{exp} = 1.9 \pm 0.4$ m s$^{-1}$.

\begin{table}
   \caption[]{Planetary parameter best-fit values obtained through a joint modeling of Keplerian signals and correlated stellar noise, using Gaussian process regression.}
          \label{tab_planpar}
          \centering
          \scriptsize
    \begin{tabular}{l l l}
             \hline
             \noalign{\smallskip}
             Jump parameter     &  prior & Best-fit value \\
             \noalign{\smallskip}
             \hline
             \noalign{\smallskip}
             $h$ [m$\,s^{-1}$] & $\mathcal{U}$(0,10) & 1.84$^{+0.25}_{-0.20}$ \\
             \noalign{\smallskip}
             $\lambda$ [days] & $\mathcal{U}$(20,150) & 36.4$^{+8.8}_{-7.6}$ \\
             \noalign{\smallskip}
             $w$ & $\mathcal{U}$(0,1) & 0.59$^{+0.17}_{-0.15}$ \\
             \noalign{\smallskip}
             $\theta$ [days] & $\mathcal{U}(40,60)$ & 46.7$^{+4.8}_{-4.3}$ \\
             \noalign{\smallskip}
             \hline
             \noalign{\smallskip}
             $\sigma_{\rm jit, HIRES-pre}$ [m$\,s^{-1}$] & $\mathcal{U}(0,5)$ & 1.48$\pm0.72$  \\
             \noalign{\smallskip}
             $\sigma_{\rm jit, HIRES-post}$ [m$\,s^{-1}$] & $\mathcal{U}(0,5)$ & 1.01$^{+0.15}_{-0.13}$ \\
             \noalign{\smallskip}
             $\sigma_{\rm jit, HARPS-N}$ [m$\,s^{-1}$] & $\mathcal{U}(0,5)$ & 1.49$^{+0.15}_{-0.14}$ \\
             \noalign{\smallskip}
             \hline
             \noalign{\smallskip}
             $\gamma_{\rm HIRES-pre}$ [m$\,s^{-1}$] & $\mathcal{U}$(-100,100) & 2.2$\pm$1.6 \\
             \noalign{\smallskip}
             $\gamma_{\rm HIRES-post}$ [m$\,s^{-1}$] & $\mathcal{U}$(-100,100) & 1.9$^{+1.1}_{-1.2}$ \\
             \noalign{\smallskip}
             $\gamma_{\rm HARPS-N}$ [m$\,s^{-1}$] & $\mathcal{U}$(-100,100) & 0.5$\pm1.1$ \\
             \noalign{\smallskip}
             \hline
             \noalign{\smallskip}
             $K_{\rm b}$ [m$\,s^{-1}$] & Modified scale invariant  & 1.68$^{+0.17}_{-0.18}$ \\
             & K$_{0}$=1; K$_{max}$=10 & \\
             \noalign{\smallskip}
             $P_{\rm b}$ [days] & $\mathcal{U}$(10.9,12.0) & 11.4407$^{+0.0017}_{-0.0016}$ \\
             \noalign{\smallskip}
             $T_{\rm 0,b}$ [BJD-2\,450\,000] & $\mathcal{U}(6994,7008)$ & 6995.86$^{+0.30}_{-0.31}$\\
             $\sqrt[]{e_\text{b}}\cdot\cos\omega_\text{b}$ & $\mathcal{U}$(-1,1) & 0.11$\pm 0.33$ \\
             \noalign{\smallskip}
             $\sqrt[]{e_\text{b}}\cdot\sin\omega_\text{b}$ & $\mathcal{U}$(-1,1) & -0.29$^{+0.43}_{-0.31}$\\
             \noalign{\smallskip}
             $K_{\rm c}$ [m$\,s^{-1}$] & Modified scale invariant  & 2.5$^{+1.3}_{-1.0}$ \\
             & K$_{0}$=1; K$_{max}$=100 & \\
             \noalign{\smallskip}
             $\ln P_{\rm c}$ [$\ln$ days] & $\mathcal{U}(\ln2000,\ln12000)$ & 8.93$^{+0.26}_{-0.25}$\\
             \noalign{\smallskip}
             & & 7600$^{+2200}_{-1700}$ [days] \\
             \noalign{\smallskip}
             $T_{\rm 0,c}$ [BJD-2\,450\,000] & $\mathcal{U}(0,12000)$ & 10730$^{+910}_{-1690}$ \\
             \noalign{\smallskip}
             $\sqrt[]{e_\text{c}}\cdot\cos\omega_\text{c}$ & $\mathcal{U}$(-1,1) & 0.10$^{+0.18}_{-0.21}$ \\
             \noalign{\smallskip}
             $\sqrt[]{e_\text{c}}\cdot\sin\omega_\text{c}$ & $\mathcal{U}$(-1,1) & 0.19$^{+0.17}_{-0.23}$\\
             \noalign{\smallskip}
             \hline
             \noalign{\smallskip} 
              $dV_{\rm r}/dt$ [m$\,s^{-1}\,day^{-1}$] & $\mathcal{U}$(-0.1,0.1) & -0.00024$^{+0.00047}_{-0.00049}$\\
             \noalign{\smallskip}
             \hline
             \noalign{\smallskip}
             \textit{Derived quantities}\tablefootmark{a} \\
             \noalign{\smallskip}
             $e_{\rm b}$ & & 0.094$^{+0.091}_{-0.065}$ \\
             \noalign{\smallskip}
             & & <0.13 ($68.3^\text{th}$ percentile) \\
             \noalign{\smallskip}
             $\omega_{\rm b}$ [rad] & & 0.96$^{+0.86}_{-1.29}$ \\
             \noalign{\smallskip}
             $M_{\rm p, b}\sini$ ($\mearth$) & & 3.03$^{+0.46}_{-0.44}$ \\
             \noalign{\smallskip} 
             $a_{\rm p, b}$ (AU) & & 0.072$^{+0.003}_{-0.004}$\\
             \noalign{\smallskip} 
             $e_{\rm c}$ & & 0.27$^{+0.28}_{-0.19}$ \\
             \noalign{\smallskip}
             & & <0.40 ($68.3^\text{th}$ percentile) \\
             \noalign{\smallskip}
             $\omega_{\rm c}$ [rad] & & -1.02$^{+1.76}_{-0.93}$ \\
             \noalign{\smallskip}
             $M_{\rm p, c}\sini$ ($\mearth$) & & 36$^{+25}_{-18}$ \\
             \noalign{\smallskip} 
             $a_{\rm p, c}$ (AU) & & 5.4$^{+1.0}_{-0.9}$\\
            \noalign{\smallskip} 
             \hline
      \end{tabular}
      \tablefoot{\tablefoottext{a}{Derived quantities from the posterior distributions. We used the following equations (assuming  $M_{\rm s}+m_{\rm p} \cong M_{\rm s}$): $M_{\rm p}\sini \cong$ ($K_{\rm p} \cdot M_{\rm s}^{\frac{2}{3}} \cdot \sqrt{1-e^{2}} \cdot P_{\rm p}^{\frac{1}{3}}) / (2\pi G)^{\frac{1}{3}}$; $a \cong [(M_{\rm s}\cdot G)^{\frac{1}{3}}\cdot P_{\rm p}^{\frac{2}{3}}]/(2\pi)^{\frac{2}{3}} $, where $G$ is the gravitational constant. }}
\end{table}

\subsection{Comparison with CARMENES results}
\label{carmenes_discussion}

In a very recent work, \citet{trifonovetal2018} have published an RV analysis at optical wavelengths of the Gl15A system as part of the CARMENES survey. The RVs derived from the visible arm of CARMENES showed no evidence of the presence of Gl15A\,b, thus the authors concluded it to be an artifact due to the stellar noise. The authors also identified a long-term trend in the combined Keck+CARMENES dataset which they proposed as due to a distant low-mass companion. As discussed in Sec. \ref{emcee_analysis}, our HARPS-N data alone clearly confirm the presence of the 11.44 d period due to Gl15A\,b, albeit with a reduced amplitude and mass, and the analysis of the combined HARPS-N+Keck datasets makes a decisive case for the existence of the long-period low-mass giant Gl15A\,c based on a 5-yr time-span of HARPS-N observations, partly overlapping with the Keck time-series. The orbital elements for Gl15A\,c derived in this work have larger formal uncertainties than those reported in \citet{trifonovetal2018} due to the likely much longer orbital period inferred for the planet, but the inferred companion mass is fully in agreement with the preliminary CARMENES estimate.

\begin{figure}
\centering
\subfloat[][]
{\includegraphics[width=0.95\hsize]{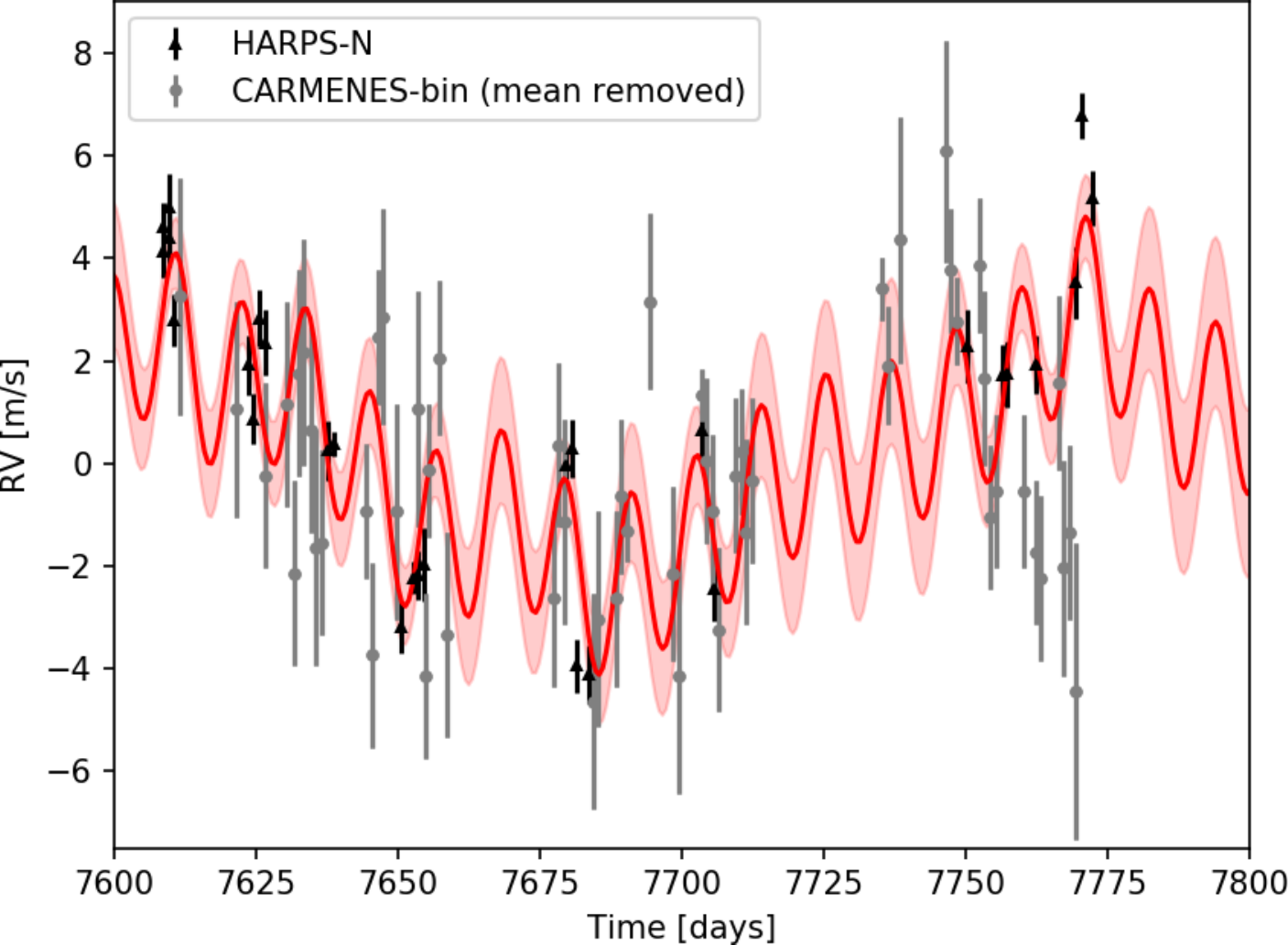}} \\
\subfloat[][]
{\includegraphics[width=0.95\hsize]{figures/periodogram_carmenes_1.pdf}}
\caption{a) Comparison between the last season HARPS-N data (black triangles) and the overlapping CARMENES data (grey dots). The red solid line represents our two-planet plus stellar noise model, with the pink shaded area representing the 1-$\sigma$ uncertainties on the GP hyper-parameters ; b) GLS periodogram of the combined HARPS-N and CARMENES RV datasets.}
\label{m22_discussion_carmenes}
\end{figure}

In Fig. \ref{m22_discussion_carmenes}a we can see the last season of HARPS-N observations, that overlaps with most of the CARMENES data, compared to our two-Keplerian plus correlated stellar noise model. This season is clearly dominated by activity-related variations in the optical HARPS-N spectra, while the internal errors for HARPS-N are typically 3 times smaller than those of CARMENES (0.6 m s$^{-1}$ vs. 1.8 m s$^{-1}$). We nonetheless tested the effect of combining the HARPS-N and CARMENES datasets: the GLS periodogram, shown in Fig. \ref{m22_discussion_carmenes}b, clearly peaks on the 11.44 d period of Gl15A\,b, and remarkably resembles Fig. \ref{fig_periodogram}, recovering the same amplitude as that recovered on the HARPS-N dataset alone.

We also tested our two planet + stellar noise model described in Sec. \ref{emcee_analysis} on the complete dataset obtained combining the HIRES, HARPS-N and CARMENES time series (running the MCMC code for around 100 times the maximum autocorrelation time), and we obtained values of the system parameters entirely in line with those presented in Table \ref{tab_planpar}. In particular, the recovered Doppler semi-amplitude and minimum mass for the inner planet are $K_b = 1.52^{+0.17}_{-0.16}$ m s$^{-1}$ and $m_b \sin i_b = 2.77^{+0.43}_{-0.40}$ M$_\oplus$ respectively. These values are well within 1-$\sigma$ of the results from the analysis of only the HIRES and HARPS-N datasets: no significant drop in the amplitude of Gl15A\,b signal is observed. It's also worth noticing that the residual jitter found in the analysis is small ($\sigma_\text{jit,CARMENES} = 0.99^{+0.32}_{-0.39}$ m s$^{-1}$) signifying that again the GP is able to model the stellar noise for this third dataset as well. The GP parameters are almost unchanged with respect to the values presented in Tab. \ref{tab_planpar} ($h = 1.77^{+0.19}_{-0.17}$ m s$^{-1}$, $\lambda = 35.5^{+8.3}_{-7.0}$ d and $\theta = 43.7^{+4.7}_{-2.6}$ d), meaning that the activity signal description derived from the first two datasets is robust also in modeling the CARMENES data. Given the intrinsically higher quality of the HARPS-N data, that drive the GP regression modeling and in which the coherence of the signal from the inner planet Gl15A\,b is clearly present, we decide to stick to the results in Table \ref{tab_planpar} for the purpose of the dynamical interaction analysis described in Sec. \ref{paperm22_ekl}.

\citet{trifonovetal2018} also stated, as an argument in favour of the non-Keplerian interpretation of Gl15A\,b, that the $11.44$ d signal disappeared when analysing the last two years of HIRES observations, subsequent to the time series used by \citet{howardetal2014}. Studying the same time span of data, we observed that, when ignoring the outlier described in Sec. \ref{rv_time_series}, the $11.44$ days signal is still clearly visible in the GLS periodogram.

\section{Binary orbital interaction}
\label{binary_interaction}

The orbital eccentricity of the outer planet, Gl15A\,c, is quite uncertain: as shown in Fig. \ref{fig_postdist}, the posterior distribution has no clear peak, only constraining the eccentricity towards small values, with a $68\%$ probability to be $<0.40$ and $95\%$ probability to be $<0.72$. This would normally point towards the adoption of a circular orbit as best fit solution for the system, lacking a significant evidence of eccentricity.

But this would be to reckon without the stellar companion.

The presence of Gl15B cannot be ignored, especially when studying a wide orbit such as that of Gl15A\,c. 

We thus investigate how the dynamical interaction with the companion star could influence the Gl15A system. 
One of the main mechanisms to excite exoplanets eccentricities is the Lidov-Kozai effect, in which the presence of an external perturber  causes oscillations of the eccentricity, $e$, and the inclination, $i$, with the same period but opposite phase.
It was originally studied by \citet{lidov1962} and \citet{kozai1962} to compute the orbits of high inclination small Solar System bodies, like asteroids and artificial satellites, and it is strongly dependent on the eccentricities and mutual inclination of the involved objects.

Thus, in order to better estimate the strength of Lidov-Kozai oscillations in the Gl15A planetary system, we need to understand as precisely as possible the orbit  of the stellar companion Gl15B.

\subsection{Orbital modeling from astrometry and RV data}
\label{orbit_mod}

The first obstacle in the dynamical analysis of the system was the poorly constrained orbital parameters of the companion.
\citet{lippincott1972} estimated a period of $2600$ yr from $\sim 100$ yr of astrometric measurements, which cover less than $4 \%$ of the orbit.

Dealing with a similar case of an eccentric planet hosted by a wide binary system, \citet{hausermarcy1999} developed a technique for constraining long-period binary orbital parameters, combining astrometric and radial velocity measurements.
The method is based on the fact that, being newtonian mechanics deterministic, knowing the instantaneous full position and velocity vector, $[x,y,z,V_x,V_y,V_z]$, of one mass with respect to the other, you can compute the exact orbit of the system.
Therefore, even by observing a small fragment of the orbit, we should be able to gather all the orbital parameters of the stellar companion.
Of course astrometry, which is restrained in the plane of the sky, cannot provide the complete 3D information needed for this analysis, so additional data from radial velocity, to compute the third component of the velocity vector, and parallax distance, to convert astrometric positions in cartesian coordinates, is necessary.

\begin{figure}
\centering
\subfloat[][]
{\includegraphics[width=0.9\hsize]{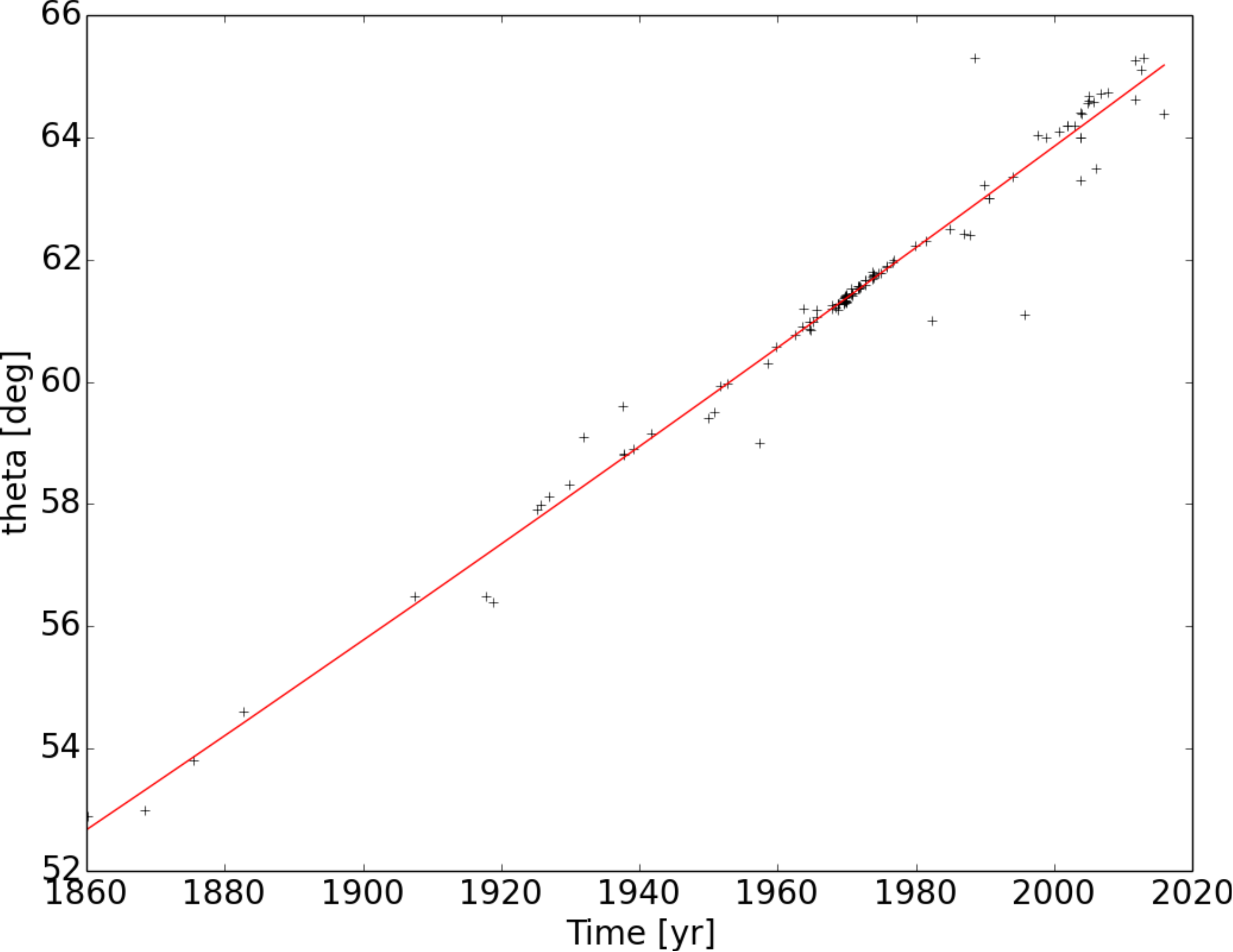}} \\
\subfloat[][]
{\includegraphics[width=0.9\hsize]{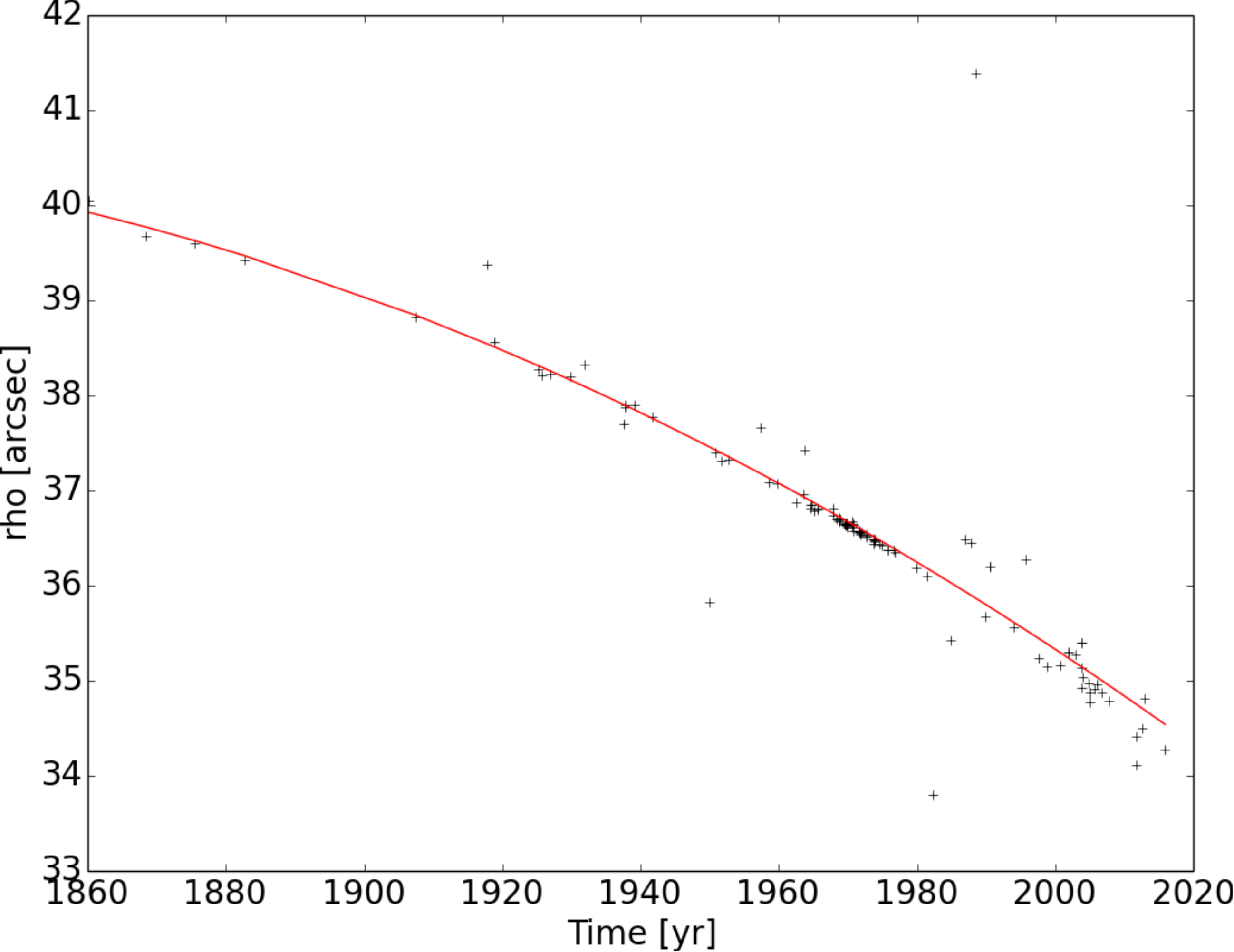}}
\caption{Position angle (a) and angular separation (b) of Gl15B with respect to Gl15A. Position angle is measured from north towards east. The red lines in both panels represent the second order polynomial fits.}
\label{astro.pos}
\end{figure}

Following the procedure of \citet{hausermarcy1999}, we downloaded 122 astrometric observations of the Gl15 system from the WDS catalogue, spanning from 1860 to 2015.
The variations of the position angle $\theta$ of Gl15B relative to Gl15A and the angular separation $\rho$ are shown in Fig. \ref{astro.pos}.
To derive $\theta$ and $\rho$ at a specific time, along with their derivatives $d \theta / dt$ and $d\rho/dt$ which we need to calculate the velocity component in the plane of the sky, we fitted the data with a second order polynomial:
\begin{equation}
\label{eq_thetafit}
 \theta_\text{fit} = a_\theta + b_\theta t + c_\theta t^2,
\end{equation}
\begin{equation}
\label{eq_rhofit}
 \rho_\text{fit} = a_\rho + b_\rho t + c_\rho t^2.
\end{equation}
From these we can easily derive $d \theta_\text{fit} / dt$ and $d\rho_\text{fit}/dt$ as:
\begin{equation}
 {d\theta_\text{fit} \over dt} = b_\theta + 2 c_\theta t,
\end{equation}
\begin{equation}
 {d\rho_\text{fit} \over dt} = b_\rho + 2 c_\rho t.
\end{equation}

Adopting the parallax value from Table \ref{star_par}, $\pi_P = 280.3 \pm 1$ mas, we can derive the Cartesian position and velocity using their Equations (1)-(4):
\begin{equation}
 x(\text{AU}) = {\rho_\text{fit} \over \pi_P} \cos \theta_\text{fit},
\end{equation}
\begin{equation}
 y(\text{AU}) = {\rho_\text{fit} \over \pi_P} \sin \theta_\text{fit},
\end{equation}
\begin{equation}
 V_x = {1 \over \pi_P} {d\rho_\text{fit} \over dt} \cos \theta_\text{fit} - {\rho_\text{fit} \over \pi_P} {d\theta_\text{fit} \over dt} \sin \theta_\text{fit},
\end{equation}
\begin{equation}
 V_y = {1 \over \pi_P} {d\rho_\text{fit} \over dt} \sin \theta_\text{fit} + {\rho_\text{fit} \over \pi_P} {d\theta_\text{fit} \over dt} \cos \theta_\text{fit},
\end{equation}
thus we gained 4 of the desired physical components, $[x,y,V_x,V_y]$, as a function of time $t$, through $\theta_\text{fit}$ and $\rho_\text{fit}$ (Eq. \ref{eq_thetafit}, \ref{eq_rhofit}).

\begin{figure}
\centering
\subfloat[][]
{\includegraphics[width=0.9\hsize]{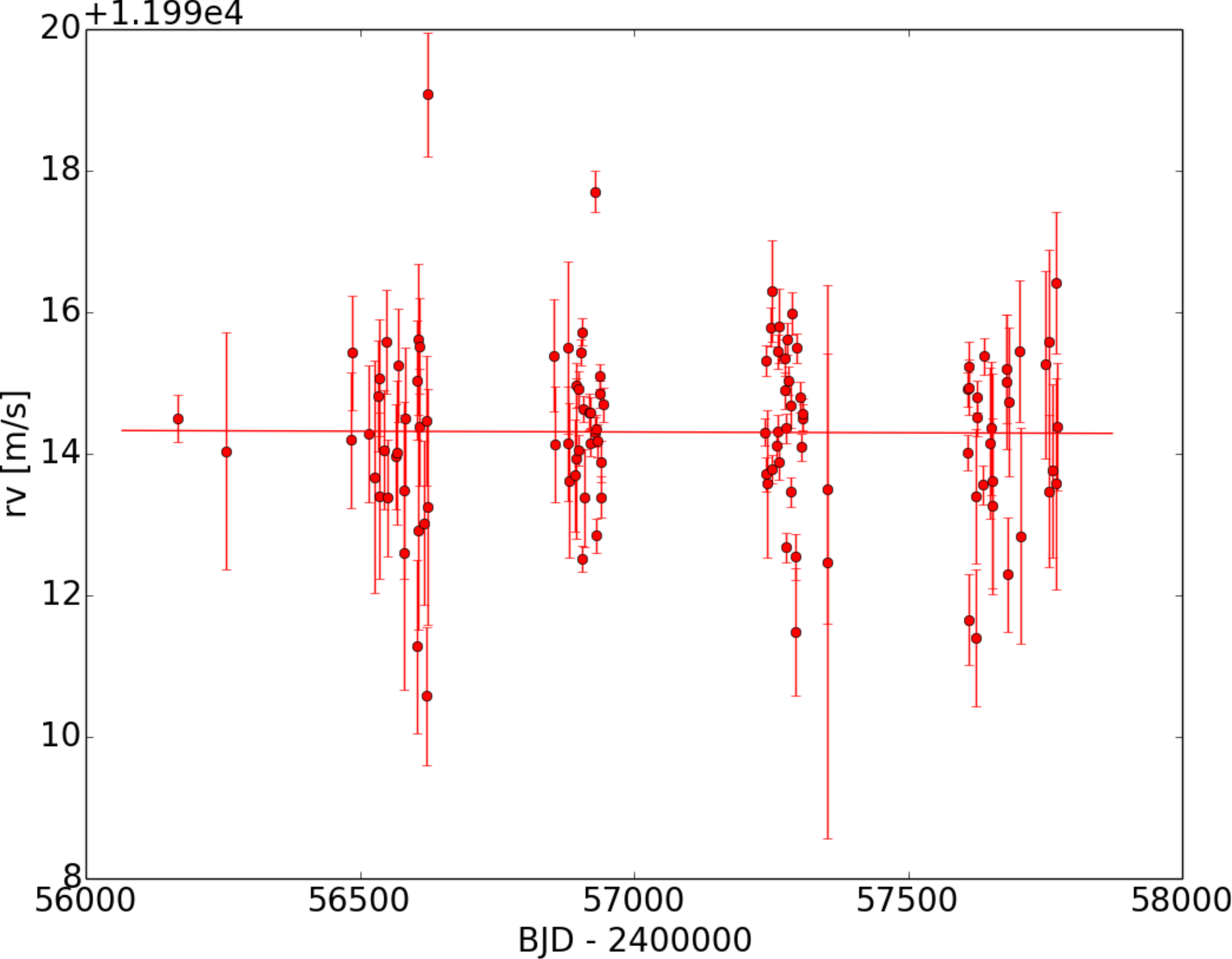}} \\
\subfloat[][]
{\includegraphics[width=0.9\hsize]{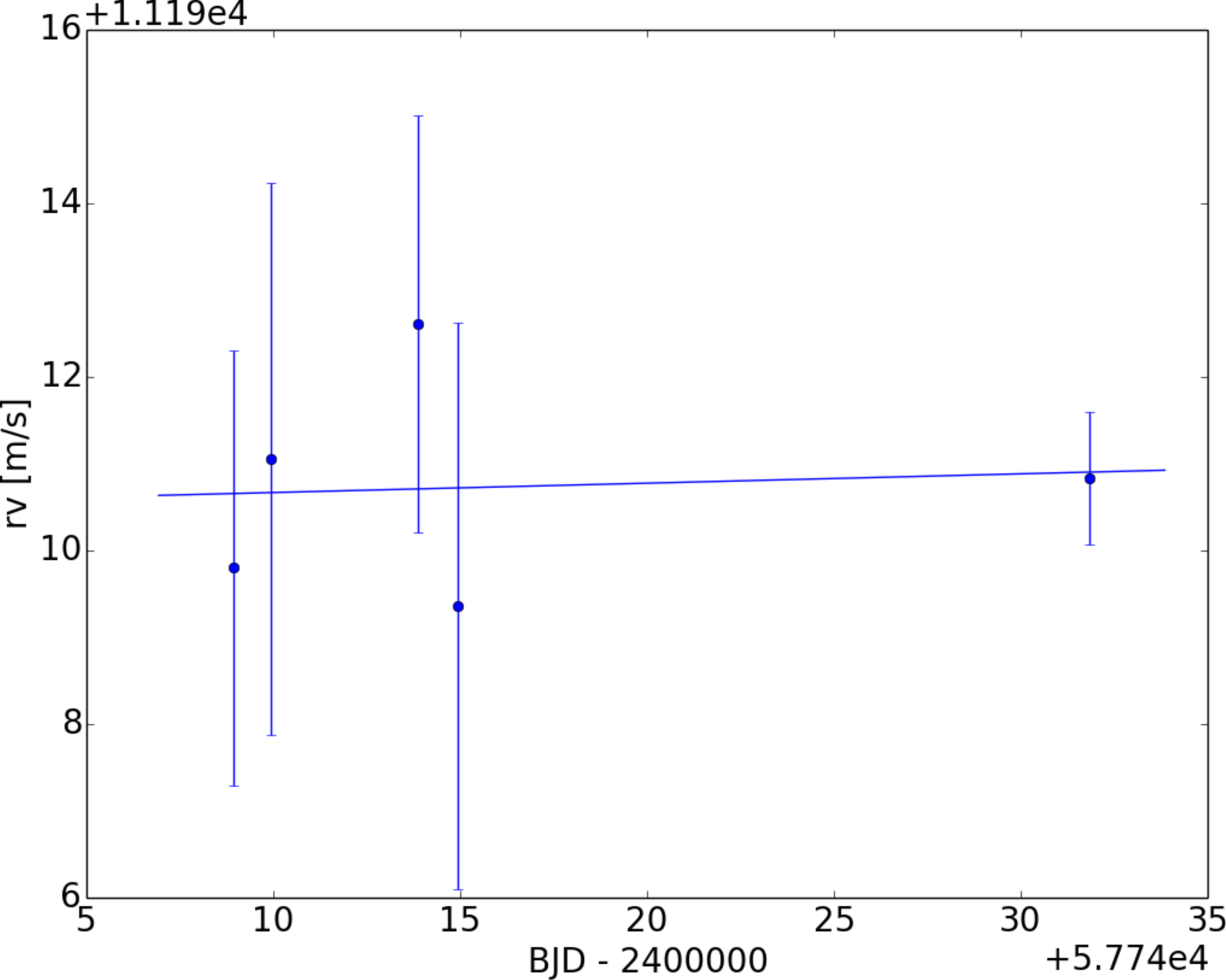}}
\caption{Radial velocities of (a) Gl15A (after subtracting the planetary and stellar signals) and (b) Gl15B. The solid lines show the respective first order polynomial fits.}
\label{abs.rv}
\end{figure}

To obtain the third component of the velocity vector, $V_z$, we use the combined Doppler information of the two stars: from the same HARPS-N spectra we used to obtain the RV time series for the planet detection, collected as illustrated in Sect. \ref{rv_time_series}, we extract the absolute radial velocities with the DRS pipeline. We use the DRS pipeline instead of TERRA since the latter only produce relative RV, which cannot be used in the comparison of two different objects.
For Gl15A we take all the 115 {HARPS-N} epochs, and subtract from the absolute RV the planetary and stellar signals, as derived in Sect. \ref{emcee_analysis}.
For Gl15B we use the 5 spectra we took on January 2017, as described in Section \ref{rv_time_series}. The two datasets are illustrated in Fig. \ref{abs.rv}. To derive the binary orbit we need to know all the position and velocity components at the same instant. Therefore we fit the two RV time series with first-order polynomials.
From the difference of the two linear fits we obtain the relative line-of-sight velocity $V_z$ of Gl15B with respect to Gl15A. We can then select an epoch, and compute the values of $[x,y,V_x,V_y,V_z]$ for that time.

The selected epoch to derive the binary orbit is BJD \nolinebreak ${= 2457754.5}$, that is January 1st 2017, which is well in between all the datasets, and close to the Gl15B RV time series, which is the shortest and most uncertain.

\begin{figure*}
\centering
\includegraphics[width=0.3\hsize]{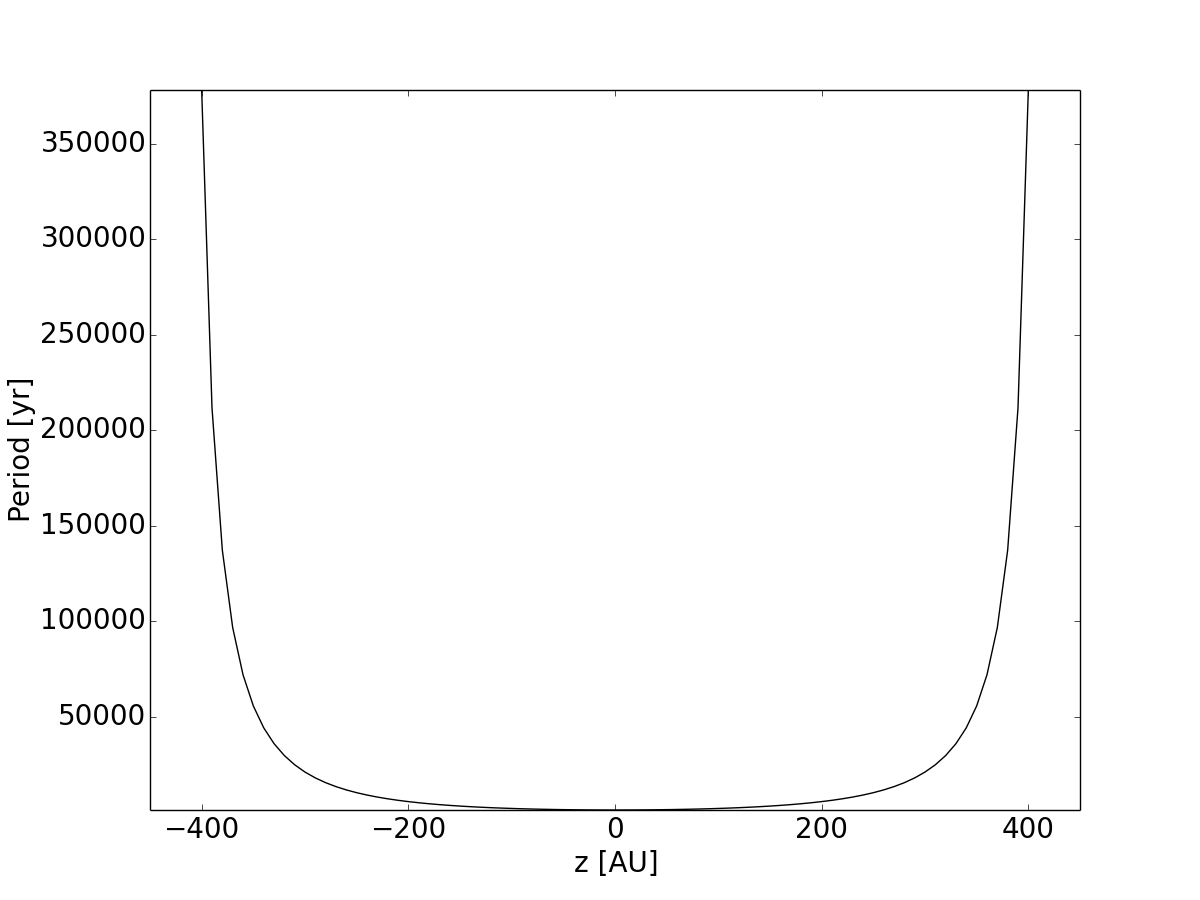} 
\includegraphics[width=0.3\hsize]{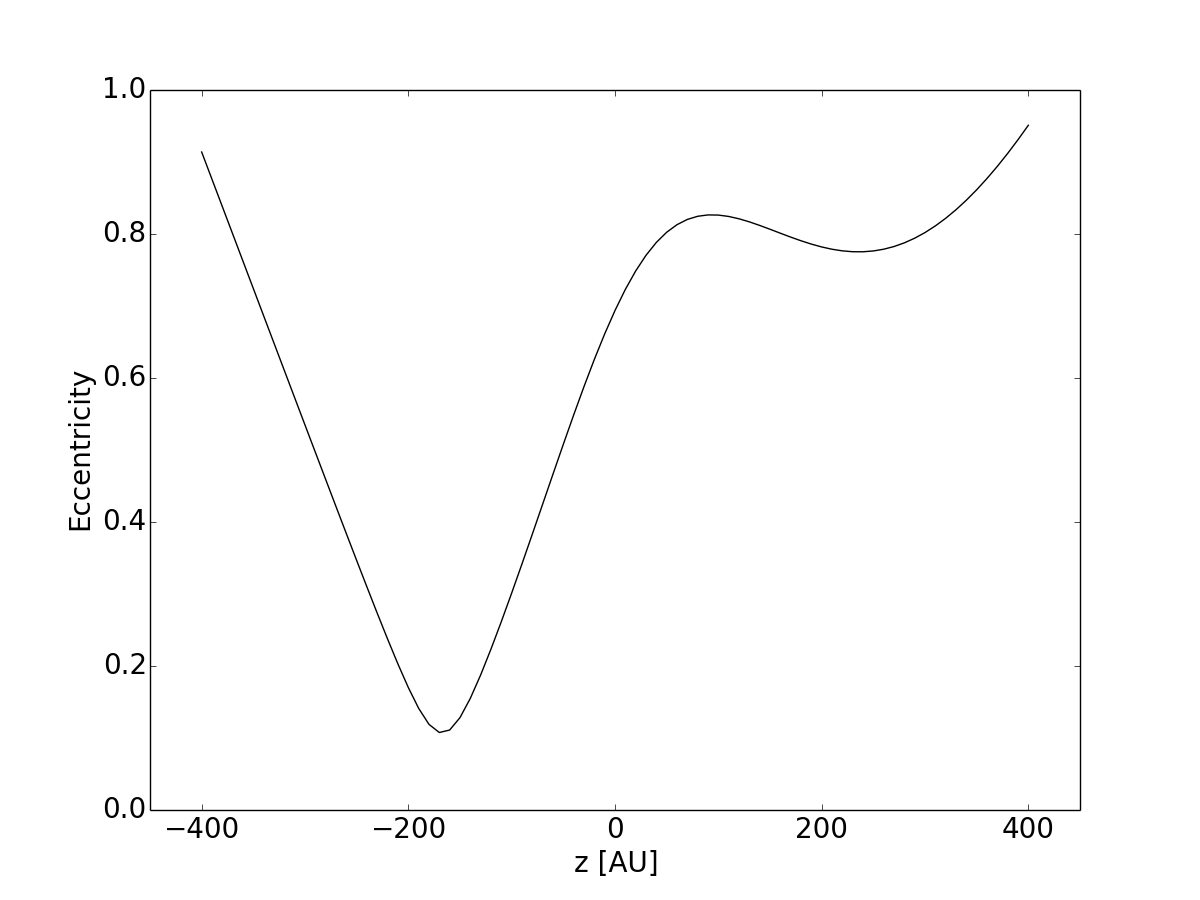} 
\includegraphics[width=0.3\hsize]{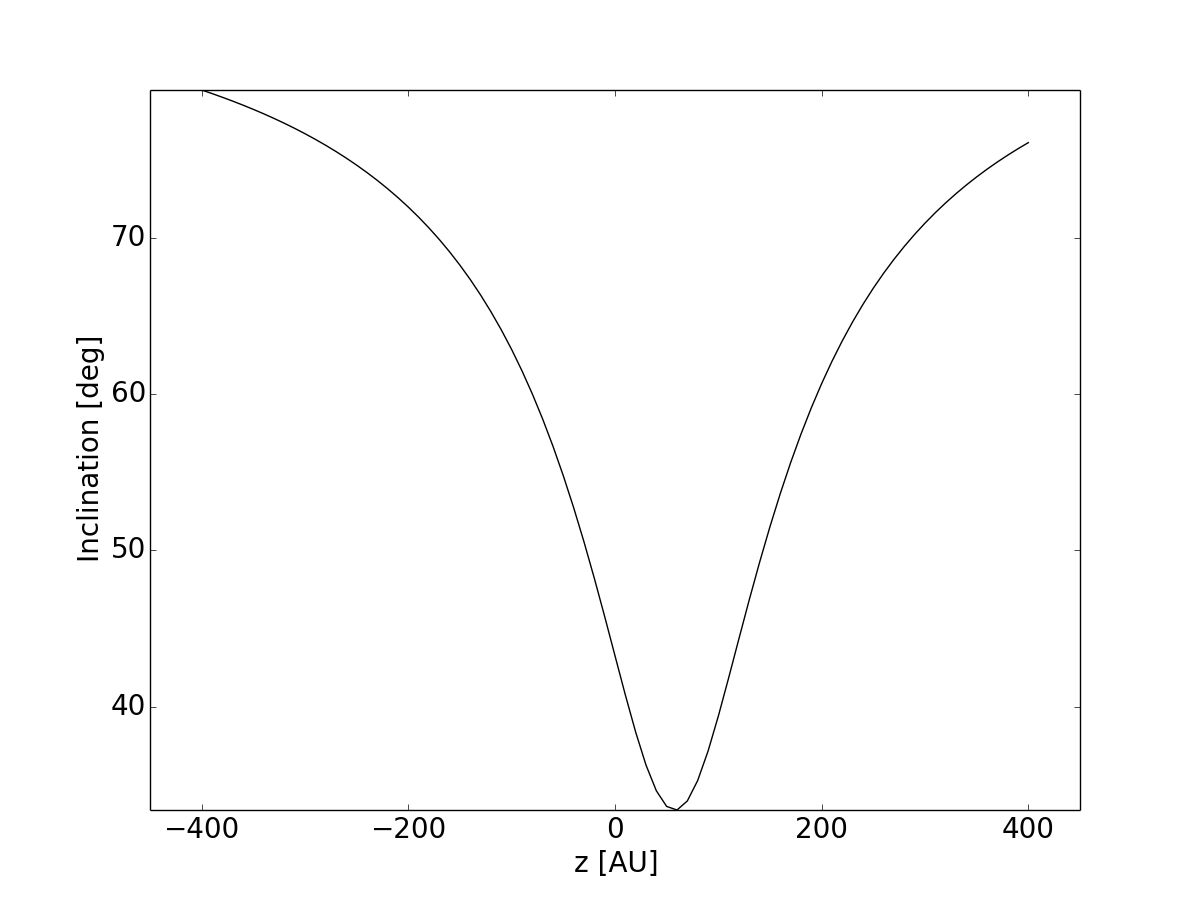} \\
\includegraphics[width=0.3\hsize]{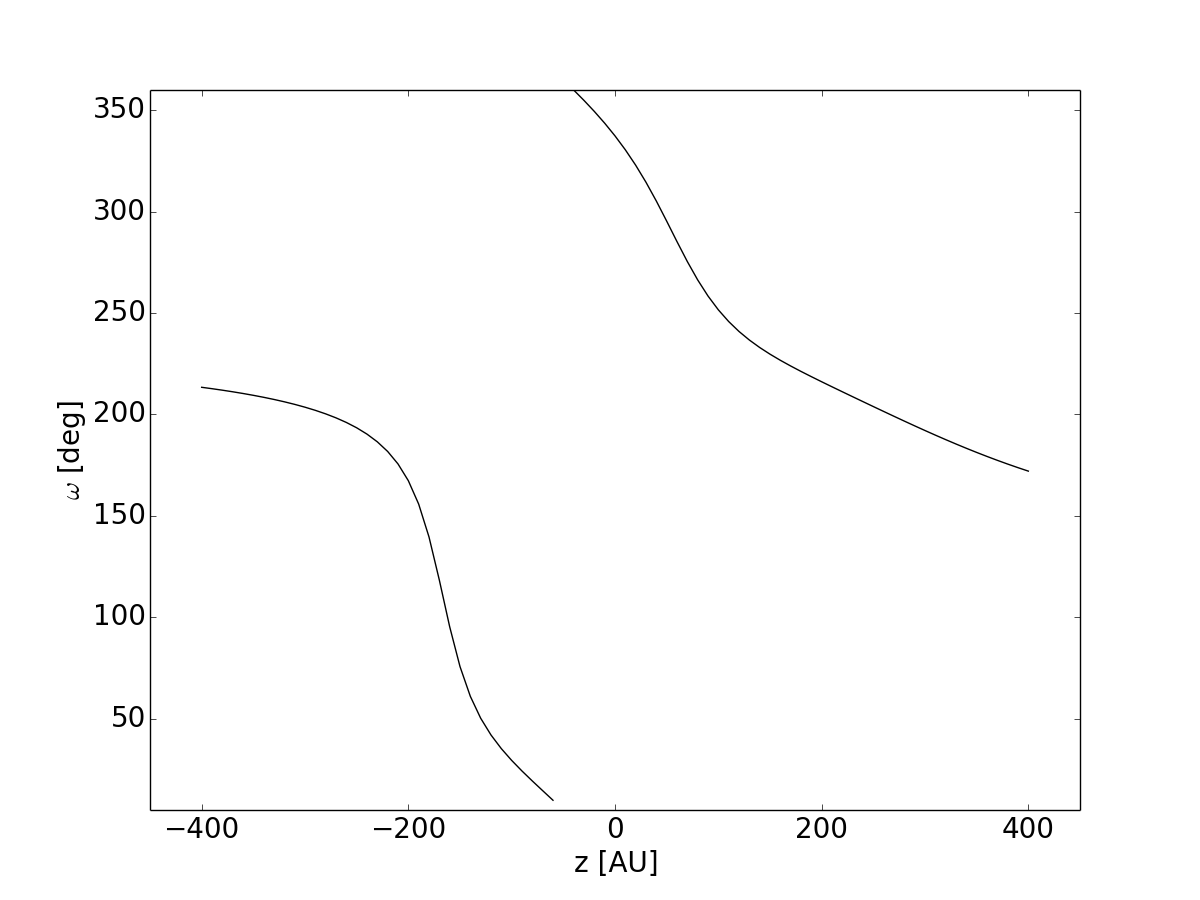} 
\includegraphics[width=0.3\hsize]{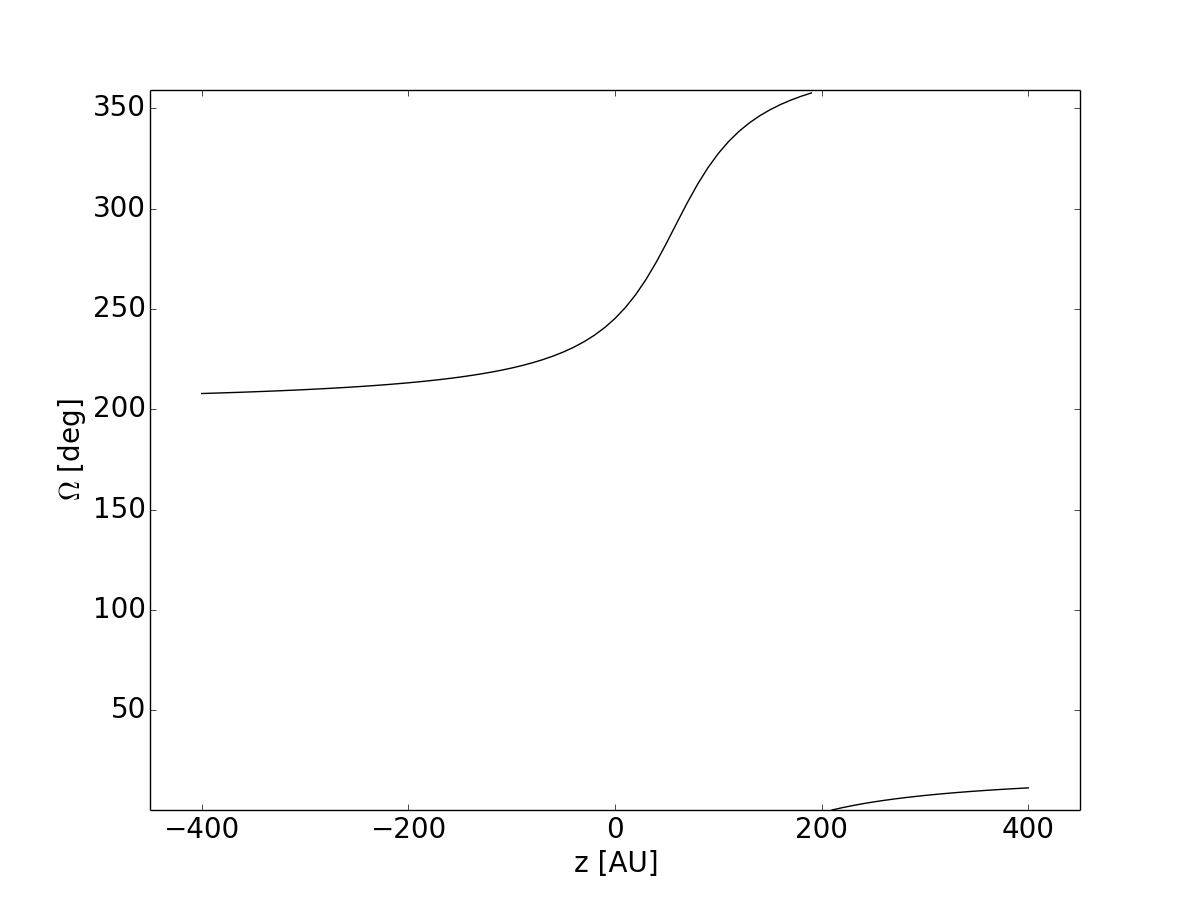} 
\includegraphics[width=0.3\hsize]{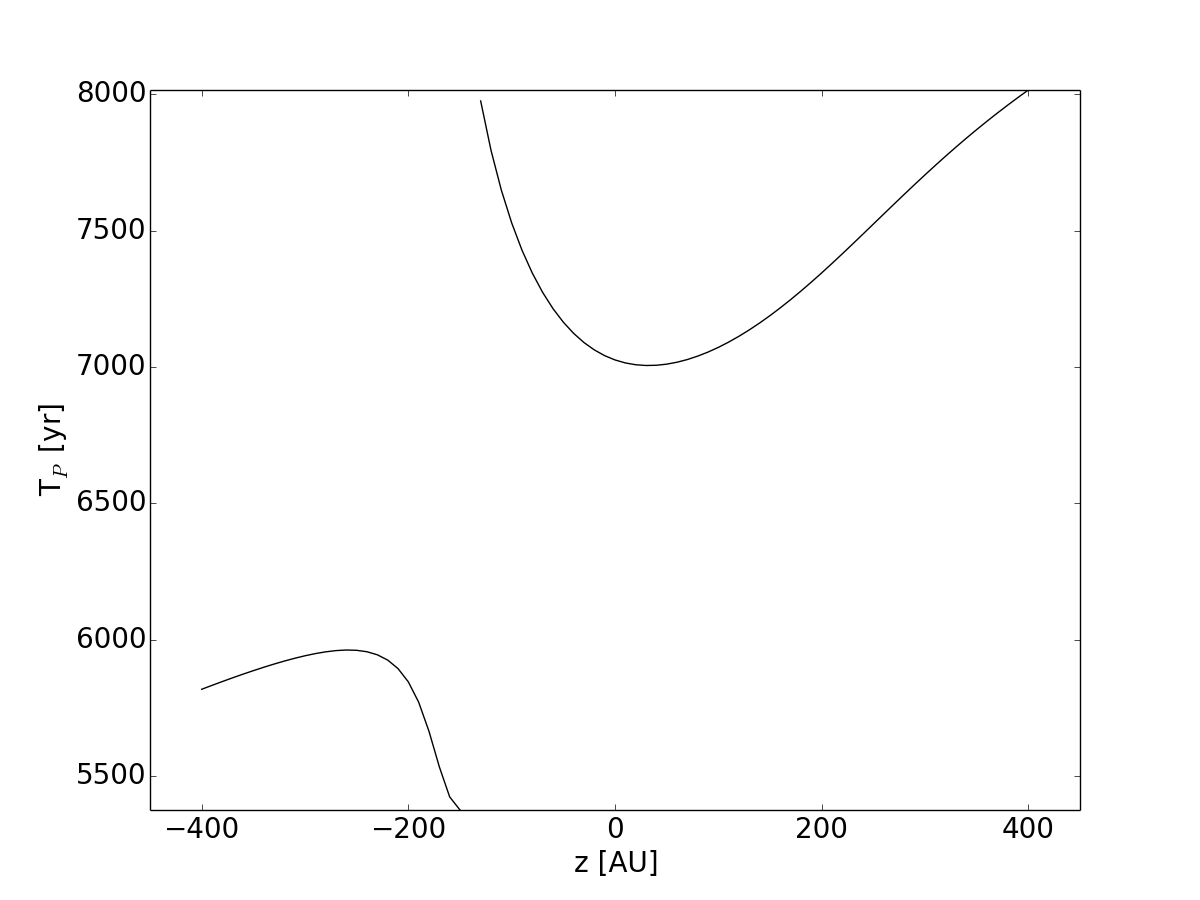}
\caption{Orbital parameters of Gl15B as a function of the line-of-sight separation $z$ with Gl15A.}
\label{zorb.res}
\end{figure*}

The last missing piece of the puzzle is, of course, the line-of-sight separation $z$, which cannot be measured, but can be constrained imposing the condition of a bound orbit for the binary system, which is expected due to the similar spectral type and proximity of the two stars. The condition of bound orbit translate into a condition on the total energy of the system:
\begin{equation}
 E = {1 \over 2} \mu v^2 - {G M_A M_B \over r} < 0,
\end{equation}
where $\mu = G (M_A+ M_B)$, $v = \sqrt{V_x^2 + V_y^2 + V_z^2}$ and, of course, $r = \sqrt{x^2 + y^2 + z^2}$.
\\From this we get a range of acceptable $z$ values $-400 \lesssim z \lesssim 400$ AU.
From every value of $z$ is possible to compute all the orbital parameters $[P,e,i,\omega,\Omega,T_P]$ for the binary system \citep[see][Eq. (7)-(15)]{hausermarcy1999}.
The results are shown in Fig. \ref{zorb.res}.
As we can see, there is a wide variety of possible orbits, with completely different eccentricities and orientations, even with the bound orbit constrain, and this is still insufficient for any meaningful dynamical analysis.

The procedure by \citet{hausermarcy1999} provides the orbital solution for every single value of $z$, but does not in any way distinguish between the more probable configurations.
But those orbital configurations are not all equally likely: from theory and observations of binary systems we know the expected distributions for different orbital parameters. From this information we can extract some priors to help us identify the 
most probable orbital configuration for the system, that is the best fit value of the line-of-sight separation $z$. 

To do this we perform a Monte Carlo simulation in which the priors on the orbital parameters are injected via rejection sampling. 
Not all the a priori distributions of the orbital parameters are known, so we restricted the prior selection to the parameters that have a strong impact on the outcome and for which a good information is available.
Due to the central role played by the eccentricity of the perturber in the Lidov-Kozai perturbation, we apply a prior on $e$, to have the best possible constraint on it.
We select the eccentricity distribution from \citet{tokovininkiyaeva2016}, who studied a sample of 477 wide binaries within 67 pc from the Sun with median projected separation of $\sim 120$ AU, very close to that of the Gl15AB system. The forementioned prior is:
\begin{equation}
 f(e) = 1.2 e + 0.4 .
\end{equation}
The second choice is a prior on the orbital period, in order to penalize long period poorly bound orbits. The chosen prior is the one suggested by \citet{duchenekraus2013} for low-mass binary stars, that is a log-normal distribution, with $\bar{a} \approx 5.2$ AU and $\sigma_{\log P} \approx 1.3$.

\begin{figure}
\centering
\includegraphics[width=0.9\hsize]{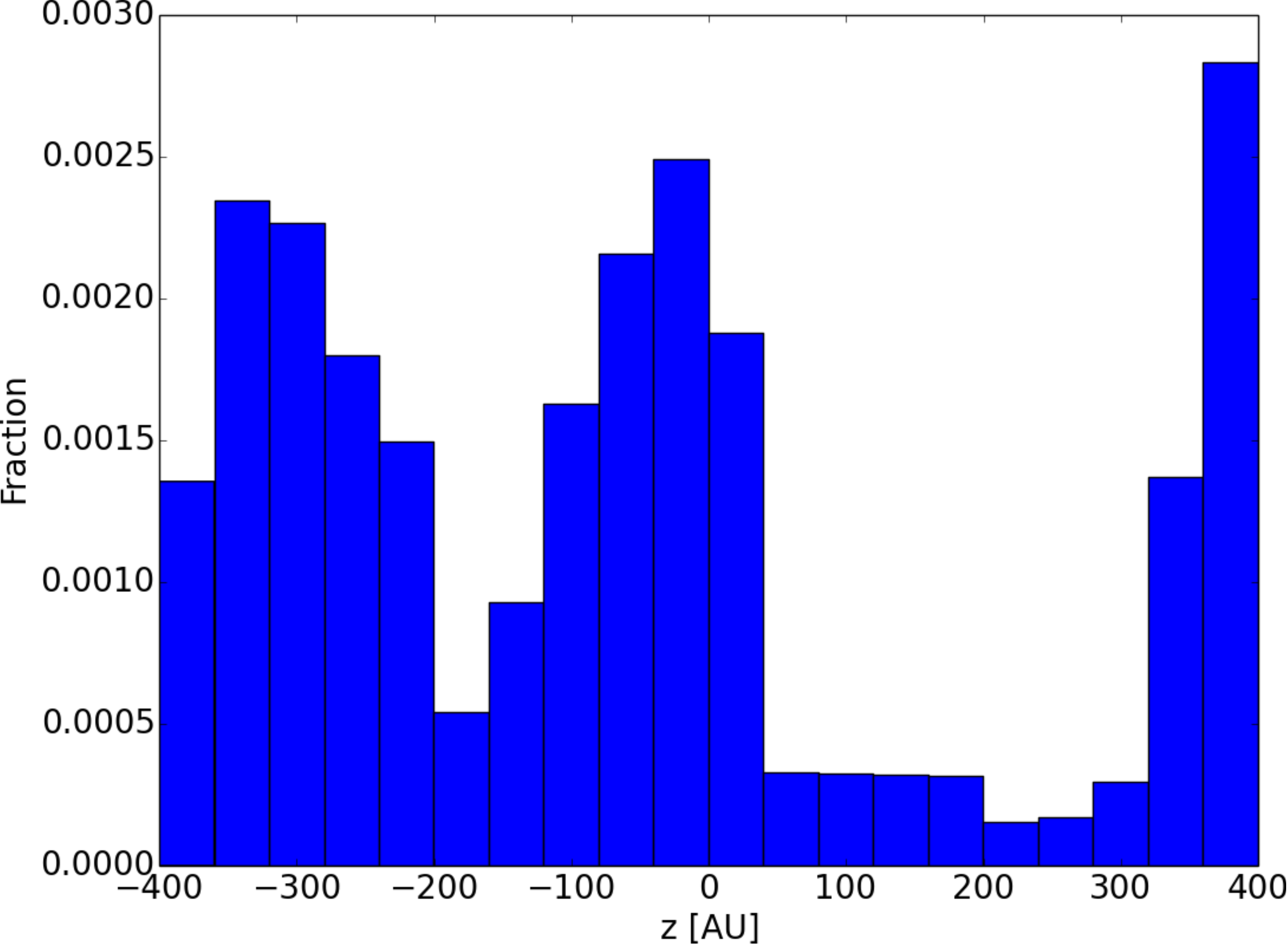}
\caption{Distribution of $z$ resulting from the Monte Carlo simulation with the $e$ and $P$ priors injected.}
\label{mc.zres}
\end{figure}

The results of the Monte Carlo simulation are illustrated in Fig. \ref{mc.zres}. As we can see there are three main peaks in the distribution. The third one, on the far right, represents orbits almost unbound, so it can be ignored, both because, as we said, the two stars are expected to be in a binary system, and because, even if bound, such wide orbits would have no influence whatsoever on the dynamics of the planetary system we intended to study. The latter can be said also on the peak on the left, which correspond to a period of $P = 22000^{+175000}_{-15000}$ yr, and a semi-major axis of $a = 640^{+2100}_{-350}$ AU, again too distant from the planetary system to have a significant influence.

We thus use the central peak of Fig. \ref{mc.zres} to derive the orbital parameters best solutions and error bars. To do this we fit a truncated normal distribution in the range $z \in [-200,200]$ AU, and use the mean value $\mu_z$ to calculate the corresponding best solution orbital parameters as described before. The upper and lower errors on the orbital parameters are calculated by taking $\mu_z+\sigma_z$ and $\mu_z-\sigma_z$ and deriving the corresponding values of $[P,e,i,\omega,\Omega,T_P]$.
The orbital parameters solutions and errors are listed in Table \ref{orb_sol}.

\begin{table}
\caption{Best orbital parameters for the Gl15 binary system from the MC simulation with priors on period and eccentricity.}             
\label{orb_sol}      
\centering                          
\begin{tabular}{ccc}        
\hline\hline                        
   \noalign{\smallskip}
   $P$ & $[$yr$]$ & $1230^{+930}_{-110}$  \\
   \noalign{\smallskip}
   $a$ & $[$AU$]$ & $93^{+42}_{-6}$ \\
   \noalign{\smallskip}
   $e$ & & $0.53^{+0.23}_{-0.28}$ \\
   \noalign{\smallskip}
   $i$ & $[$deg$]$ & $54^{+11}_{-16}$ \\
   \noalign{\smallskip}
   $\omega$ & $[$deg$]$ & $2^{+35}_{-43}$ \\
   \noalign{\smallskip}
   $\Omega$ & $[$deg$]$ & $230^{+30}_{-10}$ \\
   \noalign{\smallskip}
   $T_P$ & $[$yr$]$ & $7140^{+560}_{-140}$ \\
   \noalign{\smallskip}
\hline
\end{tabular}
\end{table}

Assuming the best fit orbital parameters listed in Table \ref{orb_sol}, we can calculate the RV signal caused by Gl15B on Gl15A, and compare it with the value of the residual acceleration found by our MCMC analysis. Gl15B RV signal is approximately linear, as is to be expected due to the small fraction of the orbit covered by the time series, and correspond to an acceleration $dV_{\rm r}/dt = -0.00024^{+0.00060}_{-0.00376}$ m s$^{-1}$ d $^{-1}$, which is fully compatible with the result from our MCMC listed in Table \ref{tab_planpar}.

\subsection{Lidov-Kozai Interaction modeling}
\label{paperm22_ekl}

The results of the previous section show that the most likely orbit of the stellar companion Gl15B has a high eccentricity, $e = 0.53$. This is a further clue that strong orbital perturbation could affect the planetary system.

The interaction mechanism studied is the Eccentric Lidov-Kozai effect (commonly reffered to with the literature-coined acronym EKL). This mechanism applies to hierarchical triple-body systems, and consists in eccentricity and inclination oscillations on timescales much larger than the orbital period of the influenced body. We can safely ignore the mutual interaction between the two planets of the Gl15A system, except in the event of close encounters, and thus treat their interaction with the binary separately, as three body systems.

The EKL mechanism is very sensitive to the mutual orientations of the orbits of the perturber, i.e. the companion star, and of the influenced body, i.e. the planet.
We have derived $i,\omega,\Omega$ of Gl15B, but we do not know either $i$ or $\Omega$ of the two planets Gl15A\,b and Gl15A\,c. Thus, some assumptions are to be made about their orbital orientation. To compare the results of the dynamical interaction model with the posterior distribution found by the MCMC analysis in Sec. \ref{emcee_analysis}, we calculate the fraction of time spent by Gl15A\,c below $e=0.40$ and $e=0.72$ ($f(e<0.40) $ and $f(e<0.72)$), which can be considered a proxy of the probabilities ($66\%$ and $95\%$, respectively) to observe the system with eccentricities under those thresholds \citep[e.g.][]{andersonlai2017}. The same can be done to study the interaction of Gl15B with Gl15A\,b, in which case the $68\%$ and $95\%$ probability levels correspond to $e=0.13$ and $e=0.25$.

We perform some preliminary tests to verify the various mechanisms which could be involved in the dynamical evolution of the system: we prove that the inner planet, Gl15A\,b, is too distant from Gl15B for any significant interaction. This is in good agreement with the extremely low level of eccentricity found in our MCMC analysis. We thus focus our efforts to study the orbit of the newly discovered Gl15A\,c; we also check for the influence of dissipative tides, which could lessen the orbital eccentricity, and find that they act on much longer timescales than the EKL mechanism, so we neglected them in our analysis.

We denote the orbital parameters of Gl15B with the subscript $_B$ and the ones of Gl15A\,c with $_c$.
All the EKL integrations were performed with a timescale of $\sim 10$ Myr.
Since we cannot rule out planet-planet scattering events to have occurred in the early phases of the system's evolution, we consider different values of the initial eccentricity $e_{c,0} = 0.0,0.1,0.2,0.3,0.4$. The other unknown is the longitude of the ascending node of the outer planet $\Omega_{c}$. However, the longitude of the ascending node influences the EKL interaction mainly by changing the relative inclination the two orbits, $\theta$, which derives from the parameters of the two orbits as:

\begin{multline}
 \theta = \arccos(\sin i_c \cos \Omega_c \sin i_B \cos \Omega_B + \\
 \sin i_c \sin \Omega_c \sin i_B \sin \Omega_B + \cos i_c \cos i_B),
\end{multline}
and thus can be ignored as long as $\theta$ is conveniently sampled, which can be obtained by changing $i_c$.
For our analysis we fix $\Omega_c = 0^\circ$ and vary the planet orbital inclination in the interval $i_c \in [-5^\circ,90^\circ]$.

\begin{figure}
\centering
\subfloat[][]
{\includegraphics[width=0.9\hsize]{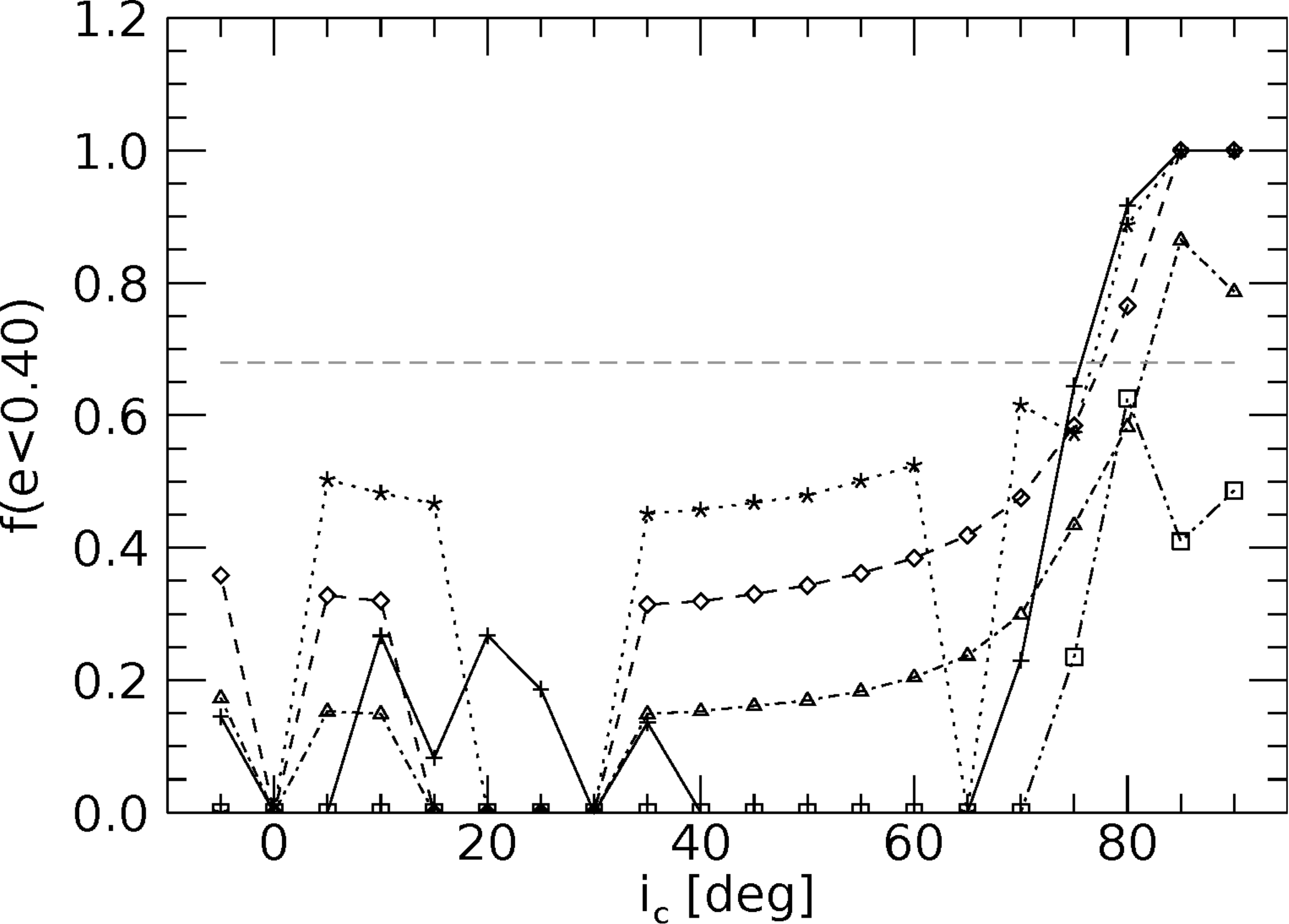}} \\
\subfloat[][]
{\includegraphics[width=0.9\hsize]{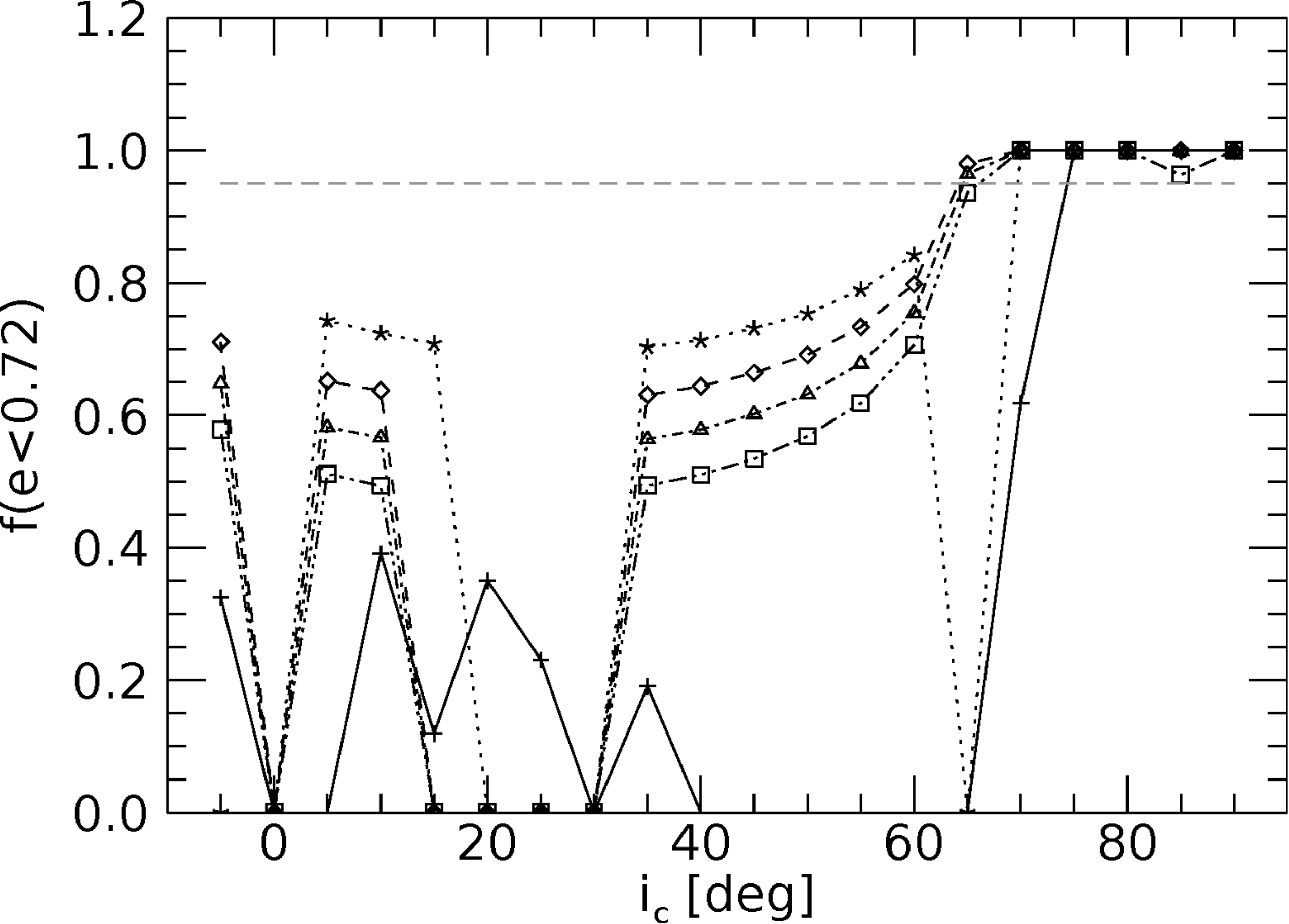}}
\caption{Fraction of time spent below the a) $e=0.40$ and b) $e=0.72$ thresholds. The different linestyles and symbols correspond to different values of $e_{c,0}$: solid and plus signs - $e_{c,0}=0$; dotted and asterisks - $e_{c,0}=0.1$; dashed and diamonds - $e_{c,0}=0.2$; dash dot and triangles - $e_{c,0}=0.3$; dash dot dot and squares $e_{c,0}=0.4$. The horizontal dashed grey lines indicate the corresponding probability from the MCMC (a) $68\%$, b) $95\%$).}
\label{ekl.36}
\end{figure}

The results are shown in Figure \ref{ekl.36}. In some cases the EKL oscillations are so extreme that the numerical integration has to be stopped due to the planet passing too close to the host star; the corresping systems are clearly unstable, and so we consider $f(e<0.40) = f(e<0.72) = 0$ to represent the incompatibility with the observed case. We also considered as unstable the cases in which the outer planet's orbit becomes too close to the inner planet's possibly producing planet-planet scattering events, that is when:
\begin{equation}
   a_c ( 1-e_{c,\text{max}}) < a_b,
\end{equation}
where $e_{c,\text{max}}$ is the maximum eccentricity reached during the EKL oscillations. In these cases we also consider $f(e<0.40) = f(e<0.72) = 0$.

As we can see in both the panels of Fig. \ref{ekl.36}, there are regions in the $i_c$ space that are unstable regardless of the initial eccentricity of the planet $e_{c,0}$: between $15^\circ$ and $30^\circ$ and around $0^\circ$. Only the solid black line, corresponding to $e_{c,0}=0$, behaves differently, showing the system to be stable between $15^\circ$ and $30^\circ$ and unstable at larger inclinations. We can also see that the Lidov-Kozai interaction is weak for $i_c \sim 90^\circ$ and strengthen as $i_c$ decreases. The top panel of Fig. \ref{ekl.36} shows that for $i_c \in [75^\circ,90^\circ]$ the resulting eccentricity ranges are compatible with the observed values. The constraints for the $e<0.72$ threshold are somewhat looser, but pointing in the same direction.

\begin{figure}
\centering
\includegraphics[width=0.9\hsize]{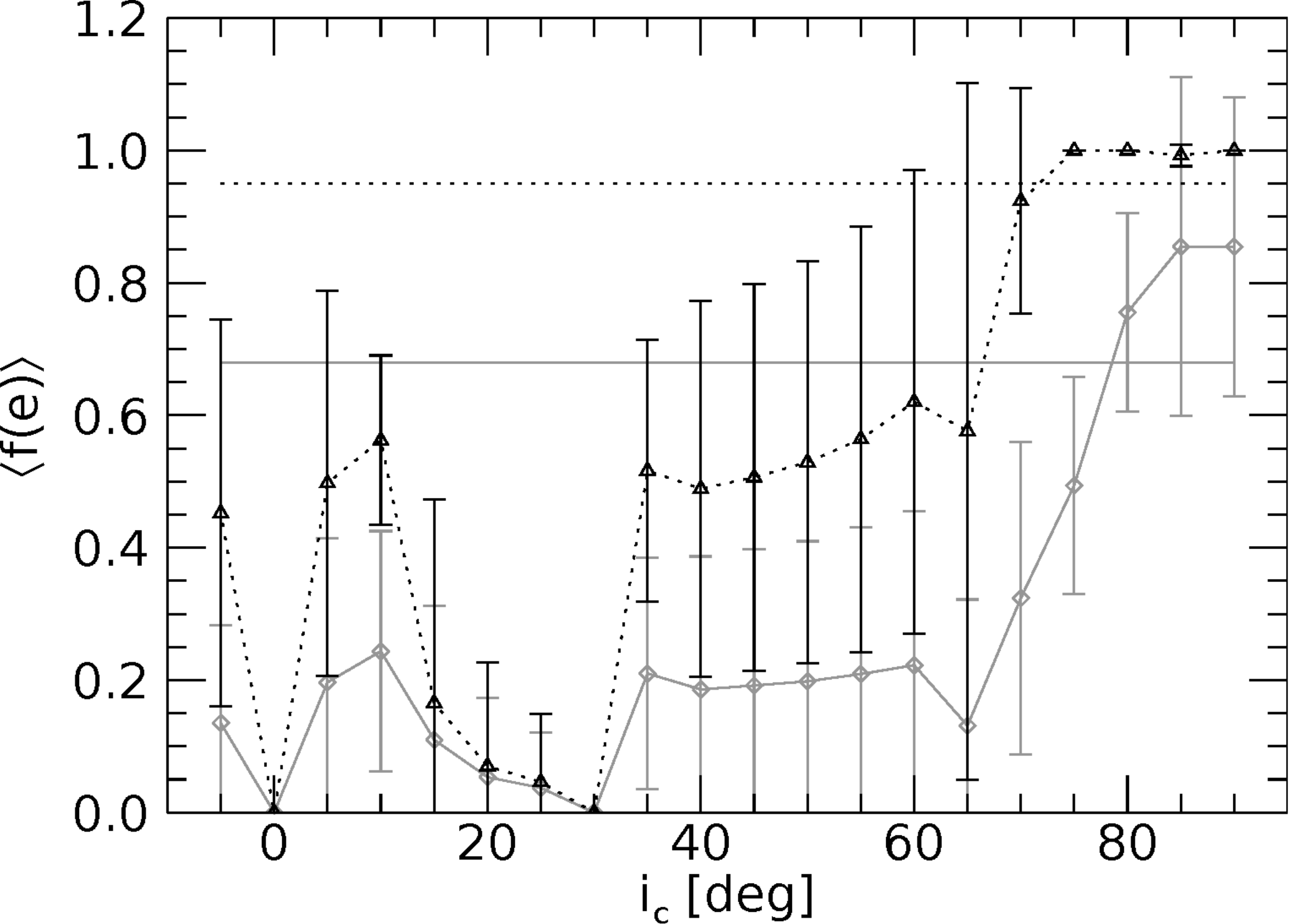}
\caption{Fraction of time spent below the $e=0.40$ (solid grey) and $e=0.72$ (dotted black) thresholds, averaged on the initial eccentricity $e_{c,0}$. The errorbars show the standard deviation.}
\label{ekl.mean}
\end{figure}

Since as previously said we do not know the initial eccentricity of the system, a more robust way to consider the dynamical evolution is to consider the average of the results of the single integrations. This can be seen in Fig. \ref{ekl.mean}, which confirms the trends just discussed, with initial inclinations around $0^\circ$ and between $15^\circ$ and $30^\circ$ leading to an unstable system, and inclinations higher than $75^\circ$ producing a good agreement with the observed orbital configuration.

\section{Discussion and conclusions}
\label{paperm22_conclusions}

We present in this paper the fifth planet detected by the HADES programme conducted with HARPS-N at TNG. This long period planet was found orbiting the planet-host M1 star Gl15A, from the analysis of high precision, high resolution RV measurements collected as part of the survey in conjunction with archive RV data from the HIRES/Keck spectrograph.

The different trends observed in the two datasets suggest the presence of a long-period companion, which is confirmed by the homogeneous Bayesian analysis of the combined RV time series. The known inner planet Gl15A\,b is also recovered. The minimum masses derived from our analysis are $M_{p,b} \sin i_b = 3.03^{+0.46}_{-0.44}$ M$_\oplus$ and $M_{p,c} \sin i_b = 36^{+25}_{-18}$ M$_\oplus$ for the inner and outer planet respectively. The mass we find for Gl15A\,b is much smaller than the value found by \citet{howardetal2014}, which was almost double due to the higher signals amplitude. The smaller value we find, can be easily explained by the additional information brought by the high-precision HARPS-N RVs, along with the new calibration of the archival HIRES data published by \citet{butleretal2017}. The combined dataset is almost twice as large that the one analysed by \citet{howardetal2014}, stretched on a significantly longer timespan, with better sampling and precision. This, together with the simultaneous modeling of the stellar activity signal, can explain the much smaller uncertainty on the minimum mass of Gl15A\,b. It also highlights the importance of taking into account the chromospheric stellar activity for the correct identification of planetary signals. It is worth noticing that instead the orbital period $P_b$ is almost unchanged from the previous estimate. Our fit also places an even lower upper limit to the eccentricity of Gl15A\,b than that found by \citet{howardetal2014} ($e < 0.13$ at a $68\%$ level of confidence), thus confirming their conclusion that the planet's motion is best described by a circular orbit.

With its period of $\simeq 21$ yr, Gl15A\,c is the longest-period sub-Jovian planet detected up to date with the RV method\footnote{\url{https://exoplanetarchive.ipac.caltech.edu/} - 28/09/2017}, the second being HD 10180 h \citep{lovisetal2011b} with a period of $\simeq 6$ yr and a minimum mass of $65.74$ M$_\oplus$ \citep{kanegelino2014}.
With the confirmed presence of two widely spaced planetary mass companions, Gl15A is now the multi-planet system closest to our Sun, at a distance of only $3.57$ pc.
 
We also compared the results of our analysis of the HIRES and HARPS-N RV data with the recent results by \citet{trifonovetal2018} based on CARMENES high-cadence monitoring of the target. \citet{trifonovetal2018} found no evidence for the presence of planet $b$, while we showed in Sec. \ref{carmenes_discussion} that the signal can be clearly detected when combining the HIRES, HARPS-N and CARMENES data, without any loss in significance. Things notwithstanding, however, the reason why CARMENES does not see the signal of Gl15A\,b is not fully clear, but the higher quality, and much longer timespan, of the HARPS-N data, combined with our modeling of the stellar activity quasi-periodic signal could be a possible explanations for the non-detection based on the CARMENES data alone. Another factor contributing to this non-detection could be the sampling of the CARMENES RV data, which appear not to be homogeneously distributed over the 11.44 d period (see Fig. 10 from \citet{trifonovetal2018}).

The CARMENES visual arm contains a spectral region extending all the way to $0.95$ $\mu$m, i.e. a significantly redder spectral range than that covered by HIRES and HARPS-N. As the amplitude of activity induced RV variations is known to be chromatic \citep{reinersetal2010}, the non-detection of the 11.44 d signal in the CARMENES time series could be an indication of a wavelength dependent-amplitude of the signal, that would clearly indicate its stellar origin. However, the CARMENES visual arm spectral range still significantly overlaps with that of HIRES and HARPS-N and thus it must be affected by stellar activity in a similar way. Based on the higher RV precision of HARPS-N, allowing a detailed modeling of quasi-periodic stellar signal (as shown in Fig. \ref{fig_starsign}), and on the coherence of the period and phase of the signal over the 20 yr time span covered by the combined HIRES and HARPS-N time series, the Keplerian origin of the signal still seems the most straightforward explanation for the observed data.
It would be however interesting to carry out a systematic study on both CARMENES and HARPS-N time series of this target, adopting the same strategy outlined by \citet{fengetal2017} for Tau Ceti, that is to study separately the RVs derived from different regions of the spectra, for a sistematic investigation of potential differences in the amplitude of the 11.44 d signal, as a function of the wavelength, but this lies well beyond the scope of this paper.

Dwarf stars are known to turn up much more frequently in multiple systems than they do in isolation, with a binary fraction as high as $\simeq 57 \%$ for Sun-like stars \citep{duqmay1991} and somewhat lower for M dwarfs \citep{bergforsetal2012}. Many young binaries possess either circumstellar or circumbinary disks \citep[e.g.][]{moninetal2007}, and the existence of stable planetary orbits in binary systems was postulated well in advance of the first exoplanets discoveries \citep{dvorak1982}.

Early studies proposed  different mass-period relations for planets around binaries and single stars \citep{zuckermazeh2002} but in the following years the evidence of such diversity decreased \citep[e.g.][]{desiderabarbieri2007,eggenberger2010}, until most recently \citet{ngoetal2017} claimed not to be any difference of planetary properties between the two kind of systems. On the other hand, recent works like \citet{moutouetal2017} find statistical evidence for a much higher binary fraction in extrasolar systems hosting eccentric exoplanets than in the ones hosting only circular planets: this points towards the confirmation of the role of stellar multiplicity in orbital excitation of planetary systems, as predicted by theoretical studies which suggested a strong orbital influence of stellar companions on planetary systems, via mechanisms such as the eccentric Lidov-Kozai (EKL) oscillations.

Our numerical analysis of the EKL effect proved the strong influence of the Gl15B on the planetary system. We show that for a narrow range of initial inclination, $75^\circ-90^\circ$, the outer planet maintains a low eccentricity orbit, regardless of the initial status in which the system was due to possible planet-planet-scattering events. We also pointed out the presence of a forbidden ranges of inclination, $15^\circ-30^\circ$ and $\sim 0^\circ$, in which the Lidov-Kozai interaction become so strong that no stable orbit can be achieved, regardless of the initial eccentricity of Gl15A\,c.

The orbital parameters of Gl15A\,c have still large uncertainties due to the observation time-span shorter than the orbital periods, and the semi-amplitude $K_c$ is significant only at a 3-$\sigma$ level, although the strong combined observational evidence from RV and imaging leaves no doubt as to the presence of a long-period planetary-mass companion. Additional RV observations in the years to come will, however, be very helpful to better constrain the orbit, and thus the mass of Gl15A\,c.

Our knowledge of this system will be also greatly improved by the results published in future Gaia data releases. For a circular orbit and assuming the minimum mass value for Gl15Ac, the expected astrometric signature on the primary is 570 $\mu$as. Gaia astrometry will only cover $\sim20\%-25\%$ of the full orbit. However, based on the \citet{torres1999} formalism curvature effects in the stellar motion should typically amount to $20-30$ $\mu$as yr$^{-2}$, thus they should be easily revealed by Gaia, that for such a bright star as Gl15A will be able to deliver end-of-mission proper motion accuracies $\lesssim10$ $\mu$as yr$^{-1}$ \citep[e.g.][]{gaiaetal2016}.

Yet even with the orbital solution now available, our analysis shows how interesting dynamical studies can be performed on the system, which, due to the presence of the eccentric binary companion, is an excellent playground to test orbital interaction mechanisms and their influence on the evolution of planetary systems.

\begin{acknowledgements}
GAPS acknowledges support from INAF through the ``Progetti Premiali'' funding scheme of the Italian Ministry of Education, University, and Research. The research leading to these results has received funding from the European Union Seventh Framework Programme (FP7/2007-2013) under Grant Agreement No. 313014 (ETAEARTH). The HARPS-N Project is a collaboration between the Astronomical Observatory of the Geneva University (lead), the CfA in Cambridge, the Universities of St. Andrews and Edinburgh, the Queen’s University of Belfast, and the TNG-INAF Observatory.\\
M.P. thanks E.T. Russo for the helpful discussions in the final stages of the analysis.
J.I.G.H., R.R.L., A.S.M. and B.T.P. acknowledge financial support from the Spanish Ministry project MINECO AYA2014-56359-P. J.I.G.H. also acknowledges financial support from the Spanish MINECO under the 2013 Ram\'on Cajal program MINECO RYC-2013-14875. A.S.M also acknowledges financial support from the  Swiss National Science Foundation (SNSF).\\
We acknowledge the computing centres of INAF – Osservatorio Astronomico di Trieste / Osservatorio Astrofisico di Catania, under the coordination of the CHIPP project, for the availability of computing resources and support.\\
This research has made use of the Washington Double Star Catalog maintained at the U.S. Naval Observatory.\\
We gratefully acknowledge an anonymous referee for her/his insightful comments that materially improved an earlier version of this manuscript.

\end{acknowledgements}

  \bibliographystyle{aa} 
  \bibliography{biblio} 

\begin{appendix}

\section{Observation log for Gl15A}

In this Section we report the observational data collected with the HARPS-N spectrograph as part the HADES project and used in the present study. We list the observation dates (barycentric Julian date or BJD), the radial velocities (RVs), and the H$\alpha$ and S$_\text{HK}$ indices, calculated by the TERRA pipeline. The RV values are corrected for perspective acceleration. The RV errors reported are the formal ones, not including the jitter term, while the H$\alpha$ and S$_\text{HK}$ errors are due to photon noise through error propagation.

\onecolumn
\begin{longtable}{ccccccc}
\caption{Data of the 115 observed HARPS-N spectra of Gl15A. We list RVs, S-index and H$\alpha$ obtained using the TERRA pipeline.}\\
\hline
\hline
BJD$-2450000$ & RV & RV$_\text{Err}$ & S$_\text{HK}$ & S$_\text{HK, Err}$ &  H$\alpha$ & H$\alpha_\text{Err}$ \\
$[d]$ & $[$m$/$s$]$ & $[$m$/$s$]$ & & \\
\hline
\endfirsthead
\caption{Continued.} \\
\hline
\hline
BJD$-2450000$ & RV & RV$_\text{Err}$ & S$_\text{HK}$ & S$_\text{HK, Err}$ &  H$\alpha$ & H$\alpha_\text{Err}$ \\
$[d]$ & $[$m$/$s$]$ & $[$m$/$s$]$ & & \\
\hline
\endhead
\hline
\endfoot
6166.848867 & $ -0.91$ & $0.68$ & $ 0.01918 $ & $  0.00018$ & $ 0.06652 $ & $  0.00007$  \\ 
6255.876480 & $ -1.80$ & $0.91$ & $ 0.01810 $ & $  0.00024$ & $ 0.06662 $ & $  0.00008$  \\ 
6484.220683 & $ -1.03$ & $0.51$ & $ 0.02151 $ & $  0.00015$ & $ 0.06747 $ & $  0.00005$  \\ 
6486.207072 & $  1.29$ & $0.37$ & $ 0.02395 $ & $  0.00013$ & $ 0.06748 $ & $  0.00005$  \\ 
6516.105891 & $ -5.77$ & $0.35$ & $ 0.01798 $ & $  0.00014$ & $ 0.06860 $ & $  0.00005$  \\ 
6527.142303 & $ -4.12$ & $0.48$ & $ 0.01823 $ & $  0.00024$ & $ 0.06749 $ & $  0.00009$  \\ 
6533.026400 & $  0.09$ & $0.41$ & $ 0.02325 $ & $  0.00013$ & $ 0.06790 $ & $  0.00005$  \\ 
6534.094838 & $ -2.28$ & $0.48$ & $ 0.02201 $ & $  0.00018$ & $ 0.06723 $ & $  0.00006$  \\ 
6535.081620 & $  1.04$ & $0.40$ & $ 0.02176 $ & $  0.00012$ & $ 0.06721 $ & $  0.00005$  \\ 
6544.187292 & $ -2.78$ & $0.48$ & $ 0.02137 $ & $  0.00013$ & $ 0.06667 $ & $  0.00005$  \\ 
6548.262049 & $  0.23$ & $0.63$ & $ 0.01938 $ & $  0.00013$ & $ 0.06677 $ & $  0.00004$  \\ 
6549.164873 & $ -5.60$ & $0.37$ & $ 0.02117 $ & $  0.00013$ & $ 0.06745 $ & $  0.00005$  \\ 
6564.110417 & $ -6.34$ & $0.39$ & $ 0.01835 $ & $  0.00011$ & $ 0.06800 $ & $  0.00005$  \\ 
6566.200787 & $ -4.31$ & $0.56$ & $ 0.01786 $ & $  0.00016$ & $ 0.06716 $ & $  0.00005$  \\ 
6569.068819 & $ -0.95$ & $0.47$ & $ 0.01965 $ & $  0.00013$ & $ 0.06727 $ & $  0.00004$  \\ 
6581.057187 & $ -2.45$ & $0.59$ & $ 0.01929 $ & $  0.00019$ & $ 0.06708 $ & $  0.00007$  \\ 
6581.156794 & $ -4.41$ & $0.73$ & $ 0.02078 $ & $  0.00032$ & $ 0.06744 $ & $  0.00010$  \\ 
6583.020532 & $ -1.89$ & $0.43$ & $ 0.01978 $ & $  0.00015$ & $ 0.06718 $ & $  0.00006$  \\ 
6602.989444 & $ -0.79$ & $0.52$ & $ 0.01804 $ & $  0.00013$ & $ 0.06651 $ & $  0.00005$  \\ 
6604.026875 & $ -3.58$ & $0.51$ & $ 0.01796 $ & $  0.00018$ & $ 0.06694 $ & $  0.00007$  \\ 
6605.008646 & $  0.66$ & $0.45$ & $ 0.01812 $ & $  0.00015$ & $ 0.06681 $ & $  0.00006$  \\ 
6605.977986 & $ -2.41$ & $0.52$ & $ 0.01877 $ & $  0.00020$ & $ 0.06640 $ & $  0.00008$  \\ 
6607.030046 & $ -2.23$ & $0.41$ & $ 0.02026 $ & $  0.00011$ & $ 0.06660 $ & $  0.00004$  \\ 
6607.998287 & $ -2.54$ & $0.49$ & $ 0.01808 $ & $  0.00013$ & $ 0.06639 $ & $  0.00004$  \\ 
6617.071400 & $ -2.93$ & $0.48$ & $ 0.02044 $ & $  0.00018$ & $ 0.06671 $ & $  0.00006$  \\ 
6620.956887 & $ -6.09$ & $0.51$ & $ 0.02112 $ & $  0.00015$ & $ 0.06697 $ & $  0.00005$  \\ 
6621.952049 & $ -2.78$ & $0.38$ & $ 0.01978 $ & $  0.00014$ & $ 0.06644 $ & $  0.00005$  \\ 
6622.846354 & $  3.57$ & $0.51$ & $ 0.02066 $ & $  0.00013$ & $ 0.06655 $ & $  0.00006$  \\ 
6622.927384 & $ -0.95$ & $0.51$ & $ 0.02059 $ & $  0.00025$ & $ 0.06657 $ & $  0.00009$  \\ 
6854.148449 & $  3.74$ & $0.42$ & $ 0.02085 $ & $  0.00013$ & $ 0.06759 $ & $  0.00005$  \\ 
6855.133137 & $  2.70$ & $0.44$ & $ 0.01961 $ & $  0.00013$ & $ 0.06712 $ & $  0.00005$  \\ 
6879.119051 & $  0.93$ & $0.41$ & $ 0.02098 $ & $  0.00013$ & $ 0.06679 $ & $  0.00006$  \\ 
6880.111019 & $  0.06$ & $0.51$ & $ 0.01927 $ & $  0.00020$ & $ 0.06654 $ & $  0.00007$  \\ 
6881.051273 & $ -2.86$ & $0.46$ & $ 0.02045 $ & $  0.00018$ & $ 0.06675 $ & $  0.00007$  \\ 
6892.106308 & $  0.03$ & $0.48$ & $ 0.02515 $ & $  0.00014$ & $ 0.06733 $ & $  0.00005$  \\ 
6893.097130 & $ -0.45$ & $0.52$ & $ 0.02380 $ & $  0.00019$ & $ 0.06692 $ & $  0.00007$  \\ 
6894.116412 & $ -0.79$ & $0.57$ & $ 0.02510 $ & $  0.00024$ & $ 0.06699 $ & $  0.00007$  \\ 
6898.093322 & $  0.00$ & $0.42$ & $ 0.02329 $ & $  0.00014$ & $ 0.06663 $ & $  0.00006$  \\ 
6899.101562 & $  1.19$ & $0.45$ & $ 0.02357 $ & $  0.00017$ & $ 0.06683 $ & $  0.00006$  \\ 
6903.116227 & $  0.59$ & $0.39$ & $ 0.02179 $ & $  0.00012$ & $ 0.06700 $ & $  0.00005$  \\ 
6904.118322 & $ -2.35$ & $0.40$ & $ 0.02097 $ & $  0.00012$ & $ 0.06718 $ & $  0.00004$  \\ 
6905.107049 & $ -0.93$ & $0.48$ & $ 0.01987 $ & $  0.00012$ & $ 0.06668 $ & $  0.00005$  \\ 
6907.101944 & $ -3.50$ & $0.46$ & $ 0.02028 $ & $  0.00012$ & $ 0.06704 $ & $  0.00004$  \\ 
6908.985370 & $ -2.61$ & $0.41$ & $ 0.01973 $ & $  0.00011$ & $ 0.06591 $ & $  0.00004$  \\ 
6918.174757 & $ -4.33$ & $0.51$ & $ 0.02241 $ & $  0.00018$ & $ 0.06647 $ & $  0.00006$  \\ 
6919.103773 & $ -4.07$ & $0.44$ & $ 0.01966 $ & $  0.00013$ & $ 0.06592 $ & $  0.00005$  \\ 
6920.111852 & $ -2.36$ & $0.41$ & $ 0.02025 $ & $  0.00013$ & $ 0.06588 $ & $  0.00005$  \\ 
6928.182581 & $  2.97$ & $0.69$ & $ 0.02386 $ & $  0.00025$ & $ 0.06589 $ & $  0.00005$  \\ 
6929.189838 & $ -1.96$ & $0.45$ & $ 0.02453 $ & $  0.00017$ & $ 0.06579 $ & $  0.00004$  \\ 
6930.172789 & $ -4.24$ & $0.54$ & $ 0.02511 $ & $  0.00018$ & $ 0.06606 $ & $  0.00006$  \\ 
6931.178229 & $ -2.74$ & $0.40$ & $ 0.02877 $ & $  0.00015$ & $ 0.06695 $ & $  0.00005$  \\ 
6932.150255 & $ -2.78$ & $0.47$ & $ 0.02413 $ & $  0.00017$ & $ 0.06616 $ & $  0.00006$  \\ 
6937.151979 & $  0.03$ & $0.40$ & $ 0.02447 $ & $  0.00012$ & $ 0.06645 $ & $  0.00004$  \\ 
6938.156285 & $ -1.45$ & $0.42$ & $ 0.02362 $ & $  0.00014$ & $ 0.06623 $ & $  0.00004$  \\ 
6939.166910 & $ -2.52$ & $0.47$ & $ 0.02437 $ & $  0.00021$ & $ 0.06616 $ & $  0.00007$  \\ 
6940.148657 & $ -2.59$ & $0.47$ & $ 0.02396 $ & $  0.00021$ & $ 0.06605 $ & $  0.00006$  \\ 
6943.154919 & $ -1.75$ & $0.41$ & $ 0.02399 $ & $  0.00018$ & $ 0.06668 $ & $  0.00006$  \\ 
7239.105162 & $ -1.71$ & $0.51$ & $ 0.01976 $ & $  0.00013$ & $ 0.06731 $ & $  0.00004$  \\ 
7240.105428 & $  0.25$ & $0.43$ & $ 0.01999 $ & $  0.00014$ & $ 0.06739 $ & $  0.00005$  \\ 
7241.110741 & $  0.22$ & $0.49$ & $ 0.02031 $ & $  0.00016$ & $ 0.06740 $ & $  0.00006$  \\ 
7242.174259 & $  1.20$ & $1.06$ & $ 0.02077 $ & $  0.00063$ & $ 0.06754 $ & $  0.00020$  \\ 
7249.150937 & $ -0.19$ & $0.58$ & $ 0.01918 $ & $  0.00018$ & $ 0.06668 $ & $  0.00006$  \\ 
7250.132604 & $  2.27$ & $0.76$ & $ 0.01974 $ & $  0.00036$ & $ 0.06689 $ & $  0.00023$  \\ 
7251.119086 & $ -1.86$ & $0.49$ & $ 0.01961 $ & $  0.00012$ & $ 0.06607 $ & $  0.00005$  \\ 
7260.085625 & $ -3.42$ & $0.51$ & $ 0.01891 $ & $  0.00014$ & $ 0.06703 $ & $  0.00005$  \\ 
7261.104456 & $ -2.98$ & $0.52$ & $ 0.01934 $ & $  0.00014$ & $ 0.06716 $ & $  0.00005$  \\ 
7262.077326 & $  0.38$ & $0.54$ & $ 0.01798 $ & $  0.00015$ & $ 0.06693 $ & $  0.00006$  \\ 
7263.077130 & $  0.37$ & $0.72$ & $ 0.01739 $ & $  0.00031$ & $ 0.06731 $ & $  0.00013$  \\ 
7264.076296 & $  2.37$ & $0.61$ & $ 0.01705 $ & $  0.00015$ & $ 0.06721 $ & $  0.00007$  \\ 
7274.063877 & $  0.18$ & $0.33$ & $ 0.01772 $ & $  0.00016$ & $ 0.06766 $ & $  0.00006$  \\ 
7275.059745 & $  1.04$ & $0.50$ & $ 0.01714 $ & $  0.00015$ & $ 0.06770 $ & $  0.00006$  \\ 
7276.086887 & $  1.85$ & $0.47$ & $ 0.01764 $ & $  0.00012$ & $ 0.06743 $ & $  0.00005$  \\ 
7277.056366 & $  2.08$ & $0.45$ & $ 0.01785 $ & $  0.00012$ & $ 0.06749 $ & $  0.00005$  \\ 
7278.081817 & $  4.50$ & $0.36$ & $ 0.02192 $ & $  0.00015$ & $ 0.06829 $ & $  0.00005$  \\ 
7282.074687 & $  3.18$ & $0.41$ & $ 0.01868 $ & $  0.00012$ & $ 0.06708 $ & $  0.00005$  \\ 
7285.103183 & $ -0.04$ & $0.58$ & $ 0.02232 $ & $  0.00022$ & $ 0.06763 $ & $  0.00008$  \\ 
7286.111204 & $  0.07$ & $0.52$ & $ 0.01900 $ & $  0.00013$ & $ 0.06683 $ & $  0.00005$  \\ 
7287.088796 & $  1.75$ & $0.40$ & $ 0.01936 $ & $  0.00019$ & $ 0.06687 $ & $  0.00007$  \\ 
7293.085637 & $  0.28$ & $0.51$ & $ 0.01998 $ & $  0.00014$ & $ 0.06666 $ & $  0.00006$  \\ 
7294.182824 & $  0.43$ & $0.28$ & $ 0.02074 $ & $  0.00023$ & $ 0.06735 $ & $  0.00007$  \\ 
7296.078102 & $  3.43$ & $0.43$ & $ 0.01962 $ & $  0.00014$ & $ 0.06684 $ & $  0.00004$  \\ 
7303.047014 & $  3.59$ & $0.56$ & $ 0.01884 $ & $  0.00013$ & $ 0.06713 $ & $  0.00005$  \\ 
7304.155243 & $  0.49$ & $0.57$ & $ 0.01856 $ & $  0.00012$ & $ 0.06719 $ & $  0.00005$  \\ 
7307.062280 & $ -1.23$ & $0.32$ & $ 0.01846 $ & $  0.00010$ & $ 0.06670 $ & $  0.00005$  \\ 
7308.066343 & $ -1.44$ & $0.50$ & $ 0.01766 $ & $  0.00013$ & $ 0.06642 $ & $  0.00006$  \\ 
7351.945509 & $ -0.58$ & $1.38$ & $ 0.01746 $ & $  0.00059$ & $ 0.06696 $ & $  0.00021$  \\ 
7352.988924 & $ -3.25$ & $0.71$ & $ 0.01699 $ & $  0.00028$ & $ 0.06705 $ & $  0.00011$  \\ 
7608.098368 & $  4.92$ & $0.48$ & $ 0.02183 $ & $  0.00016$ & $ 0.06511 $ & $  0.00005$  \\ 
7608.156042 & $  4.46$ & $0.51$ & $ 0.02098 $ & $  0.00016$ & $ 0.06486 $ & $  0.00005$  \\ 
7609.171933 & $  5.32$ & $0.66$ & $ 0.02147 $ & $  0.00026$ & $ 0.06510 $ & $  0.00009$  \\ 
7609.214653 & $  4.73$ & $0.58$ & $ 0.02157 $ & $  0.00023$ & $ 0.06509 $ & $  0.00007$  \\ 
7610.102917 & $  3.11$ & $0.51$ & $ 0.02155 $ & $  0.00043$ & $ 0.06526 $ & $  0.00014$  \\ 
7623.043877 & $  2.23$ & $0.58$ & $ 0.01970 $ & $  0.00014$ & $ 0.06586 $ & $  0.00006$  \\ 
7624.020637 & $  1.19$ & $0.50$ & $ 0.01972 $ & $  0.00015$ & $ 0.06608 $ & $  0.00006$  \\ 
7625.147350 & $  3.15$ & $0.56$ & $ 0.01994 $ & $  0.00014$ & $ 0.06574 $ & $  0.00005$  \\ 
7626.135880 & $  2.68$ & $0.64$ & $ 0.01922 $ & $  0.00017$ & $ 0.06561 $ & $  0.00006$  \\ 
7637.111528 & $  0.57$ & $0.57$ & $ 0.01936 $ & $  0.00017$ & $ 0.06671 $ & $  0.00007$  \\ 
7638.104063 & $  0.71$ & $0.24$ & $ 0.01991 $ & $  0.00017$ & $ 0.06667 $ & $  0.00006$  \\ 
7649.978472 & $ -2.87$ & $0.51$ & $ 0.01973 $ & $  0.00016$ & $ 0.06566 $ & $  0.00006$  \\ 
7652.160637 & $ -1.91$ & $0.32$ & $ 0.02100 $ & $  0.00015$ & $ 0.06494 $ & $  0.00005$  \\ 
7653.096944 & $ -1.86$ & $0.47$ & $ 0.02042 $ & $  0.00020$ & $ 0.06588 $ & $  0.00006$  \\ 
7654.113113 & $ -1.65$ & $0.72$ & $ 0.02028 $ & $  0.00024$ & $ 0.06549 $ & $  0.00008$  \\ 
7678.940556 & $  0.27$ & $0.26$ & $ 0.02201 $ & $  0.00013$ & $ 0.06649 $ & $  0.00004$  \\ 
7680.012720 & $  0.60$ & $0.58$ & $ 0.01955 $ & $  0.00013$ & $ 0.06645 $ & $  0.00006$  \\ 
7681.051921 & $ -3.64$ & $0.53$ & $ 0.01985 $ & $  0.00012$ & $ 0.06605 $ & $  0.00005$  \\ 
7683.000891 & $ -3.81$ & $0.56$ & $ 0.01973 $ & $  0.00016$ & $ 0.06642 $ & $  0.00006$  \\ 
7702.871273 & $  0.97$ & $0.54$ & $ 0.02013 $ & $  0.00016$ & $ 0.06577 $ & $  0.00006$  \\ 
7705.083646 & $ -2.15$ & $0.61$ & $ 0.02194 $ & $  0.00027$ & $ 0.06561 $ & $  0.00008$  \\ 
7749.873391 & $  2.60$ & $0.72$ & $ 0.02302 $ & $  0.00021$ & $ 0.06577 $ & $  0.00008$  \\ 
7755.850174 & $  2.03$ & $0.60$ & $ 0.01999 $ & $  0.00020$ & $ 0.06544 $ & $  0.00007$  \\ 
7756.879954 & $  2.04$ & $0.64$ & $ 0.01947 $ & $  0.00016$ & $ 0.06536 $ & $  0.00006$  \\ 
7761.876227 & $  2.22$ & $0.56$ & $ 0.01790 $ & $  0.00018$ & $ 0.06559 $ & $  0.00006$  \\ 
7768.832396 & $  3.83$ & $0.69$ & $ 0.01794 $ & $  0.00023$ & $ 0.06556 $ & $  0.00007$  \\ 
7769.842847 & $  7.07$ & $0.45$ & $ 0.01800 $ & $  0.00015$ & $ 0.06604 $ & $  0.00006$  \\ 
7771.869016 & $  5.47$ & $0.54$ & $ 0.01792 $ & $  0.00014$ & $ 0.06574 $ & $  0.00005$  \\              
\end{longtable}
\twocolumn

\end{appendix}

\end{document}